\PassOptionsToPackage{unicode}{hyperref}
\PassOptionsToPackage{hyphens}{url}
\PassOptionsToPackage{dvipsnames,svgnames,x11names}{xcolor}
\documentclass[
  12pt]{article}

\usepackage{amsmath,amssymb}
\usepackage{algorithm}
\usepackage{algorithmic}
\usepackage{iftex}
\ifPDFTeX
  \usepackage[T1]{fontenc}
  \usepackage[utf8]{inputenc}
  \usepackage{textcomp} % provide euro and other symbols
\else % if luatex or xetex
  \usepackage{unicode-math}
  \defaultfontfeatures{Scale=MatchLowercase}
  \defaultfontfeatures[\rmfamily]{Ligatures=TeX,Scale=1}
\fi
\usepackage{lmodern}
\ifPDFTeX\else  
\fi
\IfFileExists{upquote.sty}{\usepackage{upquote}}{}
\IfFileExists{microtype.sty}{% use microtype if available
  \usepackage[]{microtype}
  \UseMicrotypeSet[protrusion]{basicmath} % disable protrusion for tt fonts
}{}
\makeatletter
\@ifundefined{KOMAClassName}{% if non-KOMA class
  \IfFileExists{parskip.sty}{%
    \usepackage{parskip}
  }{% else
    \setlength{\parindent}{0pt}
    \setlength{\parskip}{6pt plus 2pt minus 1pt}}
}{% if KOMA class
  \KOMAoptions{parskip=half}}
\makeatother
\makeatletter
\ifx\paragraph\undefined\else
  \let\oldparagraph\paragraph
  \renewcommand{\paragraph}{
    \@ifstar
      \xxxParagraphStar
      \xxxParagraphNoStar
  }
  \newcommand{\xxxParagraphStar}[1]{\oldparagraph*{#1}\mbox{}}
  \newcommand{\xxxParagraphNoStar}[1]{\oldparagraph{#1}\mbox{}}
\fi
\ifx\subparagraph\undefined\else
  \let\oldsubparagraph\subparagraph
  \renewcommand{\subparagraph}{
    \@ifstar
      \xxxSubParagraphStar
      \xxxSubParagraphNoStar
  }
  \newcommand{\xxxSubParagraphStar}[1]{\oldsubparagraph*{#1}\mbox{}}
  \newcommand{\xxxSubParagraphNoStar}[1]{\oldsubparagraph{#1}\mbox{}}
\fi
\makeatother

\usepackage{longtable,booktabs,array}
\usepackage{calc} % for calculating minipage widths
\usepackage{etoolbox}
\makeatletter
\patchcmd\longtable{\par}{\if@noskipsec\mbox{}\fi\par}{}{}
\makeatother
\IfFileExists{footnotehyper.sty}{\usepackage{footnotehyper}}{\usepackage{footnote}}
\makesavenoteenv{longtable}
\usepackage{graphicx}
\makeatletter
\def\maxwidth{\ifdim\Gin@nat@width>\linewidth\linewidth\else\Gin@nat@width\fi}
\def\maxheight{\ifdim\Gin@nat@height>\textheight\textheight\else\Gin@nat@height\fi}
\makeatother
\setkeys{Gin}{width=\maxwidth,height=\maxheight,keepaspectratio}
\makeatletter
\def\fps@figure{htbp}
\makeatother

\makeatletter
\@ifpackageloaded{caption}{}{\usepackage{caption}}
\AtBeginDocument{%
\ifdefined\contentsname
  \renewcommand*\contentsname{Table of contents}
\else
  \newcommand\contentsname{Table of contents}
\fi
\ifdefined\listfigurename
  \renewcommand*\listfigurename{List of Figures}
\else
  \newcommand\listfigurename{List of Figures}
\fi
\ifdefined\listtablename
  \renewcommand*\listtablename{List of Tables}
\else
  \newcommand\listtablename{List of Tables}
\fi
\ifdefined\figurename
  \renewcommand*\figurename{Figure}
\else
  \newcommand\figurename{Figure}
\fi
\ifdefined\tablename
  \renewcommand*\tablename{Table}
\else
  \newcommand\tablename{Table}
\fi
}
\@ifpackageloaded{float}{}{\usepackage{float}}
\floatstyle{ruled}
\@ifundefined{c@chapter}{\newfloat{codelisting}{h}{lop}}{\newfloat{codelisting}{h}{lop}[chapter]}
\floatname{codelisting}{Listing}

\makeatother
\makeatletter
\@ifpackageloaded{caption}{}{\usepackage{caption}}
\@ifpackageloaded{subcaption}{}{\usepackage{subcaption}}
\makeatother

\ifLuaTeX
  \usepackage{selnolig}  % disable illegal ligatures
\fi
\usepackage[]{natbib}
\usepackage{bookmark}

\IfFileExists{xurl.sty}{\usepackage{xurl}}{} % add URL line breaks if available
\hypersetup{
  pdftitle={Title},
  pdfauthor={Author 1; Author 2},
  pdfkeywords={3 to 6 keywords, that do not appear in the title},
  colorlinks=true,
  linkcolor={blue},
  filecolor={Maroon},
  citecolor={Blue},
  urlcolor={Blue},
  pdfcreator={LaTeX via pandoc}}

\newcommand{\anon}{1}

\usepackage{bm}
\usepackage{amsthm}
\usepackage{amsmath}
\usepackage{bbm}
\usepackage{comment}
\newtheorem{theorem}{Theorem}

\newtheorem{remark}{Remark}[section]

\newtheorem{example}{Example}
\newtheorem{proposition}{Proposition}[section]
\newtheorem{corollary}{Corollary}[section]
\usepackage{pgffor}
\RequirePackage{natbib}
\usepackage{natbib}
\usepackage{multirow}
\def\boxit#1{\vbox{\hrule\hbox{\vrule\kern6pt\vbox{\kern6pt#1\kern6pt}\kern6pt\vrule}\hrule}}

\usepackage{xr}
\usepackage[usenames,dvipsnames,svgnames,table]{xcolor}

\providecommand{\norm}[1]{\left\|#1\right\|}

\newcommand*{\var}{\textnormal{var}}

\newcommand{\balpha}{\bm{\alpha}}
\newcommand{\bbeta}{\bm{\beta}}

\newcommand{\btheta}{\bm{\theta}}

\newcommand{\bTheta}{\bm{\Theta}}

\newcommand{\bSigma}{\bm{\Sigma}}

\newcommand{\bOmega}{\bm{\Omega}}

\newcommand{\argmin}{\mathop{\mathrm{argmin}}}

\newcommand \btt{\bbeta}
\newcommand \bta{\btheta}
\newcommand \hbt{\hat{\btt}}

\newcommand \hbta{\hat{\bta}}

\def\##1\#{\begin{align}#1\end{align}}
\def\$#1\${\begin{align*}#1\end{align*}}

\usepackage{relsize}
\newcommand{\T}{{\mathsmaller {\rm T}}}

\newcommand{\BB}{\mathbb{B}}

\newcommand{\Hc}{\mathcal{H}}

\newcommand{\RR}{\mathbb{R}}
\newcommand{\PP}{\mathbb{P}}
\newcommand{\EE}{\mathbb{E}}
\newcommand{\bB}{\bm{B}}

\newcommand{\bI}{\bm{I}}

\newcommand{\bX}{\bm{X}}

\newcommand{\br}{\bm{r}}

\newcommand{\bfsym}[1]{\ensuremath{\boldsymbol{#1}}}
\def \balpha   {\bfsym{\alpha}}       \def \bbeta    {\bfsym{\beta}}

\newcommand{\Rom}[1]{\text{\uppercase\expandafter{\romannumeral #1\relax}}}

\usepackage{geometry}
\usepackage{setspace}
\usepackage[shortlabels]{enumitem}

\numberwithin{equation}{section}
\newcommand{\ES}{{\rm ES}}

\begin{document}

\def\spacingset#1{\renewcommand{\baselinestretch}%
{#1}\small\normalsize} \spacingset{1}

%%%%%%%%%%%%%%%%%%%%%%%%%%%%%%%%%%%%%%%%%%%%%%%%%%%%%%%%%%%%%%%%%%%%%%%%%%%%%%

\if1\anon
{
  \title{\bf Estimation and Inference for the Joint Autoregressive Quantile-Expected Shortfall Model}
  \author{Peiyao Cai\hspace{.2cm}\\
    Department of Statistics, University of Michigan, Ann Arbor\\
and\\
Kean Ming Tan\\
 Department of Statistics, University of Michigan, Ann Arbor\\
 and\\
    Xuming He\\
    Department of Statistics and Data Science,  \\Washington University in St. Louis}
  \maketitle
} \fi

\if0\anon
{
  \bigskip
  \bigskip
  \bigskip
  \begin{center}
    {\LARGE\bf Estimation and Inference for the Joint Autoregressive Quantile-Expected Shortfall Model}
\end{center}
  \medskip
} \fi

\bigskip
\begin{abstract}
Expected shortfall is defined as the truncated mean of a random variable that falls below a specified quantile level and is widely recognized as an important risk measure. 
    Motivated by the empirical observation of clustering patterns in financial risks, we consider a joint autoregressive model for both conditional quantile and expected shortfall. 
    Existing estimation methods for such models typically rely on minimizing a nonlinear and nonconvex joint loss function, which is challenging to solve. 
    We introduce a two-step regression framework to estimate the model parameters with theoretical guarantees.
    We further develop a Portmanteau test for model diagnostics and prove its asymptotic validity for Type I error control.
    Our empirical results on stock market data indicate that the proposed models effectively capture the clustering patterns and leverage effects on the conditional expected shortfall.
\end{abstract}

\noindent%
\vfill

\newpage
\spacingset{1.8} % DON'T change the spacing!

\section{Introduction}\label{sec-intro}

Financial risk management has been a prominent research area for decades.
In recent years, the expected shortfall has gained considerable attention as a risk measure in finance \citep{T2002}. Notably, the fundamental review of the trading book, an integral part of Basel III reforms, announced its adoption of the expected shortfall to replace the Value-at-Risk as the standard risk measure for regulatory supervision in the year 2016 \citep{B2016}. 
Let $Y_t$ be the return of a financial asset over some time horizon and let $\Hc_t$ be an information set available by time $t$. 
For a given quantile level $\tau \in (0,1)$, let $Q_\tau(Y_t\mid \Hc_t) = \inf\{r\in \RR; \PP(Y_t\leq r\mid \Hc_t)\geq \tau\}$ be the $\tau$-level conditional quantile of $Y_t$ given the information set $\Hc_t$, which is also referred to as the Value-at-Risk.  
Then, the $\tau$-level conditional expected shortfall (ES) of $Y_t$ can be defined as
\begin{align}\label{ESDef}
&\ES_\tau(Y_t\mid \Hc_t) = \EE\{Y_t\mid Y_t\leq Q_\tau(Y_t\mid \Hc_t) ,\Hc_t\} =  \frac{1}{\tau}\int_0^\tau Q_u(Y_t\mid \Hc_t)du.
\end{align}

Equation \eqref{ESDef} can be interpreted as the expected return of an asset, given that its return is less than or equal to the $\tau$-level conditional quantile.
Unlike the Value-at-Risk that focuses on a specific quantile level, the conditional expected shortfall aggregates the tail information of $Y_t$ beyond the conditional quantile, providing a more comprehensive representation of the tail risk. 
%This characteristic makes conditional expected shortfall an effective statistic for modeling the tail risk of financial assets.
%
% there does not exist a loss function such that the ES is the minimizer of the expected loss. 

Although the importance of the conditional expected shortfall is well recognized in different fields such as finance \citep{L2019}, operations research \citep{R2014}, and insurance \citep{BO2017}, its estimation and statistical inference remain challenging tasks.
This is primarily due to the non-elicitability of the expected shortfall, that is, there does not exist a loss function for which the minimizer of the expected loss is the expected shortfall \citep{G2011}. 
Although the expected shortfall is not elicitable by itself, it is jointly elicitable with the quantile from a class of strictly joint loss functions \citep{FZ2016}.
Motivated by the joint loss function in \cite{FZ2016}, \cite{DB2019} and \cite{P2019} proposed joint semi-parametric regression frameworks to estimate the $\tau$-level conditional quantile and expected shortfall simultaneously.
On the other hand, \cite{B2020} introduced a two-step estimation approach by first estimating the conditional quantile via quantile regression, and then estimating the expected shortfall using a least squares regression with a constructed pseudo-response based on the estimated quantile function obtained from the first step.  
Motivated by \cite{B2020}, \cite{H2023} proposed estimation and inference procedures for linear expected shortfall models under the heavy-tailed data and the growing dimension regime for the covariates. Furthermore, \cite{BL2023} and \cite{Z2023} proposed an $\ell_1$-penalized expected shortfall regression for the high-dimensional regime in which the dimension of the covariates can be larger than the number of samples.
\cite{L2022} proposed an expected shortfall regression approach based on the optimization perspective considered in \cite{R2014}.
%the estimation and inference for $l_1$-penalized expected shortfall estimators, respectively.
%Under the high dimensional linear model regime,

In this manuscript, we aim to address the following issues in the context of modeling conditional expected shortfall for financial time series:
(i) develop a model that effectively captures the risk patterns of financial markets;
(ii) provide an easy-to-implement estimation procedure with theoretical guarantees; and
(iii) construct a statistical test to assess the quality of the resulting estimator.
We start with modeling the conditional expected shortfall with autoregressive structures, i.e., the current expected shortfall can be expressed as a parametric function of the past expected shortfalls and a set of covariates. 
We are motivated by the observed clustering pattern of market risks, where a bad trading day can increase the risk level of future trading days, reflecting an important empirical property of the global financial market \citep{B1992}.
The existing literature on other risk measures such as standard deviation and Value-at-Risk with autoregressive structures is rich.
The standard deviation models include the Generalized AutoRegressive Conditional Heteroskedasticity (GARCH) and its variants \citep{B1986, L1996}.
The model proposed in \cite{E2004} assumed certain autoregressive structures on the conditional quantile, which has been extensively applied in financial risk management and prediction \citep{D2006}, and various extensions have been developed \citep{W2015}.

However, limited work has been done on estimating and conducting model diagnostics for conditional autoregressive expected shortfall in the presence of dependent data, despite its importance as a risk measure that aggregates tail risk.
Most existing work does not model the autoregressive structures of the conditional expected shortfall directly.
For instance, \citet{T2019} proposed a joint estimation approach in which they modeled the conditional quantile via \citet{E2004}, and then assumed certain parametric structures on the ratio (multiplicative models) or the difference (additive models) between the conditional quantile and the conditional expected shortfall.
A sequence of works, including \cite{G2020}, \cite{WC2023}, and \cite{P2019}, extends \citet{T2019} from the perspectives of Bayesian approaches and the generalized autoregressive score framework \citep{C2013}.
The aforementioned works have several limitations that hinder their practical applicability to financial institutions. First, the joint loss function applied is highly non-linear and non-convex, making the resulting optimization problem computationally challenging to solve in practice. Second, the estimators obtained from this joint loss function are generally not asymptotically efficient, which limits their statistical performance.
Moreover, even when estimation is feasible, a reliable diagnostic framework is important to assess model adequacy to meet the regulatory requirements for financial institutions. 
However, research on model diagnostics for expected shortfall remains relatively unexplored, with existing approaches largely relying on the joint elicitability of quantile and expected shortfall, as in \cite{NF2017}, \cite{D2017}, and \cite{BK2023}.

This manuscript contributes to the growing financial econometrics literature on the use of expected shortfall in risk management in three key aspects.
First, we propose to model the autoregressive conditional expected shortfall directly. Compared with existing literature \citep{T2019, G2020}, our modeling framework offers greater flexibilities in learning the tail risk dynamics.
We develop a computationally efficient two-step procedure to estimate the parameters of the expected shortfall models with theoretical guarantees.
 At the first stage, we solve an autoregressive conditional quantile regression problem as described in \cite{E2004}; and at the second stage, we solve a least squares autoregressive regression problem with a surrogate response variable that depends on the autoregressive conditional quantile function obtained from the first stage.
We establish the asymptotic properties of the proposed estimators and show that, owing to the orthogonality structure of our framework (see Sections~\ref{S3}--\ref{S4} for details), the quantile estimation error from the first stage is first-order negligible.
Second, we develop a novel Portmanteau test to assess the goodness-of-fit of the expected shortfall model, and we establish a theoretical guarantee for Type I error control. We further demonstrate empirically that the proposed test exhibits higher power than existing approaches in Section~\ref{S5}. Lastly, in the data application presented in Section~\ref{S6}, we uncover a leverage effect in the expected shortfall of the U.S. stock market, i.e., the expected shortfall is more sensitive to previous losses than previous gains.
In addition, motivated by the optimal weighted least squares scheme considered in \cite{B2020}, we construct a weighted least squares estimator for the proposed autoregressive expected shortfall model and show that this weighted estimator is more efficient than estimators obtained by existing approaches \citep{P2019, DB2019}, i.e., our proposed estimator has a smaller asymptotic variance.

The manuscript is organized as follows. Section \ref{S2} introduces the proposed joint autoregressive quantile-expected shortfall framework. Section \ref{S3} details the estimation and inference methodology. Section \ref{S4} presents the theoretical results for the expected shortfall estimator. Section \ref{S5} reports the results of the numerical experiments. Section \ref{S6} encompasses the empirical data analysis. We end with a discussion in Section \ref{S7}.

\section{A Joint Autoregressive Quantile-ES Model}\label{S2}
%In this section, we provide an introduction to the joint autoregressive quantile-expected shortfall framework.
%We start by defining some notation that will be used throughout the manuscript.

Consider a time series $\{Y_t, \bX_t \}_{t\geq 1}$ with $Y_t$ as the response variable and $\bX_t$ as the covariate vector at time $t$, respectively.
We allow the covariate vector $\bX_t$ to be endogenous or exogenous. In the endogenous case, $\bX_t$ consists of lagged values of $Y_t$, i.e.,  $\{Y_{t-i}\}_{i\geq 1}$.  In the exogenous case, $\bX_t$ includes additional covariates in addition to the lagged values of $Y_t$.
%The prevalent approach in time series analysis \citep{F2004,T2005} usually considers $\bX_t$ as lagged values of $Y_t$.
%\cite{H2012} and \cite{HK2014} considered the exogeneous $\bX_t$ scenarios.
For a given quantile level $\tau \in (0,1)$,
let $q_1$ and $e_1$ be the $\tau$-level (conditional) quantile and the expected shortfall of $Y_1$, respectively.
Moreover, let $\Hc_t$ be the sigma-algebra generated by $\{ q_1, e_1, Y_1, \bX_1, \ldots, Y_{t-1}, \bX_{t-1},\bX_t\}$ for any $t>1$.
%We define the conditional $\tau$-level quantile and expected shortfall of $Y_t$ correspondingly as  $Q_{\tau}(Y_t|\Hc_t) = \inf \{ r \in \mathbb{R};~P(Y_t \leq r|\Hc_t ) \geq \tau \}$ and $\ES_\tau(Y_t|\Hc_t ) = \EE\{Y_t| Y_t \leq Q_\tau (Y_t|\Hc_t), \Hc_t \}$.
Let $q_{t\tau}(\btt_\tau^*)$ and $e_{t\tau}(\bta_\tau^*)$ be two $\Hc_t$-measurable functions with true underlying parameters $\btt^*_\tau$ and $\bta^*_\tau$.
We model the conditional quantile and the ES defined in~\eqref{ESDef} as
 \begin{equation}\label{full_model}
 Q_\tau(Y_t\mid \Hc_t) = q_{t\tau}(\btt_\tau^*) \qquad \mathrm{and} \qquad \ES_\tau(Y_t\mid\Hc_t) = e_{t\tau}(\bta_\tau^*).
 \end{equation}
To simplify the notation, we omit the dependence of $q_{t\tau}(\btt_\tau^*)$ and $e_{t\tau}(\bta_\tau^*)$ on $\tau$ in the rest of the paper.
Our general model specification is motivated by empirical observations that risky days in the stock market tend to cluster together and financial risk changes smoothly over time \citep{E2004}.
Hence, it is natural to use certain autoregressive specifications to model such characteristics.
To this end, we assume that
both $q_t(\btt^*)$ and $e_t(\bta^*)$ have the following autoregressive structures:
%A generic form of our framework can be written as
 \begin{align} \label{var_es}
 q_t(\btt^*) &= \beta_0^* + \sum_{i = 1}^{l_1}\beta_i^*q_{t-i}(\btt^*) + \sum_{i = 0}^{r_1}\beta_{l_1+1+i}^* g(\bX_{t-i});\notag\\
 e_t(\bta^*) &= \theta_0^* + \sum_{i = 1}^{l_2} \theta^*_ie_{t-i}(\bta^*) + \sum_{i = 0}^{r_2}\theta_{l_2+1+i}^*h(\bX_{t-i}),
 \end{align}
 where $\{l_1, l_2, r_1, r_2 \}$ are user-specified autoregressive orders.
For notational simplicity, let $d_q = l_1 + r_1 + 2$ and $d_e = l_2 + r_2 + 2$. 
We have $\btt^* = (\beta_0^*, \beta_1^*, \ldots, \beta_{l_1 + r_1+1}^*)^\T \in \mathbb{R}^{d_q}$ and $\bta^* = (\theta_0^*, \theta_1^*, \ldots, \theta_{l_2+r_2+1}^*)^\T \in \mathbb{R}^{d_e}$.
The functions $g(\cdot)$ and $h(\cdot)$ are pre-determined and used to model relationships between $q_t(\btt^*)$ and $e_t(\bta^*)$ and the covariates $\bX_{t}$.
The selection of autoregressive orders and the functional forms of $g(\cdot)$ and $h(\cdot)$ in \eqref{var_es} depends on specific financial questions.
In practice, it is crucial to select the appropriate model specifications for \(g(\cdot)\) and \(h(\cdot)\).
%provide statistical guarantees for the model specification validities.
In Section \ref{LB}, we propose a joint quantile-expected shortfall backtesting procedure that allows us to perform statistical tests to assess model fit and determine the appropriate model specification.
In the following, we outline three special cases of~\eqref{var_es}.

\begin{example}[Symmetric Absolute Value Model]\label{SAVExp} For financial data with short-range dependencies, an autoregressive order of one is typically preferred.
We set autoregressive orders to $l_1 = l_2 =r_1 = r_2 = 1$.
When data exhibit symmetric sensitivities toward stock price impacts, both extremely high/low past returns have the same influence on the current risk level. 
To capture this pattern, we define the functions $ g(x) = h(x) = |x| $ and let
\[
q_t(\btt^*) = \beta_0^* + \beta_1^*q_{t-1}(\btt^*) + \beta_2^*|Y_{t-1}|,~~e_t(\bta^*) = \theta_0^* + \theta_1^*e_{t-1}(\bta^*) + \theta_2^*|Y_{t-1}|.
\]
\end{example}

\begin{example}[Asymmetric Slope Model]\label{ASExp}
Numerous studies have shown that the current risk level exhibits different sensitivities to past losses compared to past gains. This phenomenon is referred to as the leverage effect \citep{B1976,C1982,Yu2005}.
To model such an effect, setting $g(y) = h(y) = \{y^+, y^- \}$ yields
\begin{align*}
&q_t(\btt^*) = \beta_0^* + \beta_1^*q_{t-1}(\btt^*) + \beta_2^*Y_{t-1}^+ - \beta_3^*Y_{t-1}^- + \beta_4^*Y_{t-2}^+ - \beta_5^*Y_{t-2}^-,\\
&e_t(\bta^*) = \theta_0^* + \theta_1^*e_{t-1}(\bta^*) + \theta_2^*Y_{t-1}^+ - \theta_3^*Y_{t-1}^- + \theta_4^*Y_{t-2}^+ - \theta_5^*Y_{t-2}^-,
\end{align*}
where $y^{+}$ and $y^-$ correspond to $y\mathbbm{1}\{y\geq 0\}$  and $y \mathbbm{1}\{y< 0\}$, respectively, and $\mathbbm{1}\{\cdot\}$ is the indicator function.
\end{example}
\begin{example}[GARCH Model]\label{GARCHExp}
    Consider $Y_t = \sigma_t\eta_t$, where $\{\eta_t\}_{t\geq 1}$ is a sequence of independent and identically distributed standard normal random variables and
    $
    \sigma_t^2 = \alpha_0^* + \alpha_1^*\sigma_{t-1}^2 + \alpha_2^*Y_{t-1}^2.
    $
    This model has been widely employed to analyze the autoregressive structure in the conditional volatilities of $Y_t $.
    For $\tau < 0.5$, the conditional quantile and expected shortfall functions are
    \[
    q_t(\btt^*) = -\sqrt{\beta_0^* + \beta_1^*q_{t-1}^2(\btt^*) + \beta_2^*Y_{t-1}^2},~~e_t(\bta^*) = -\sqrt{\theta_0^* + \theta_1^*e_{t-1}^2(\bta^*) + \theta_2^*Y_{t-1}^2}.
    \]
\end{example}

\begin{remark}
   The expected shortfall, as defined in~\eqref{ESDef}, is a statistic that integrates tail information beyond the given quantile level $\tau$. The two separate model specifications in~\eqref{var_es} reflect the fact that the conditional expected shortfall may exhibit different dynamic behavior from the conditional quantile function, due to the additional tail information not captured by the quantile.
   To illustrate, consider
   \[
   Y_t = q_t\left\{\btt(U_t) \right\} = \beta_0(U_t) + \beta_1^*q_{t-1}\left\{ \btt(U_t)\right\} + \beta_2(U_t)|Y_{t-1}| + \{\beta_3(U_t) - \beta_3(\tau_0)\}|Y_{t-2}|.
   \]
   Here, $\{U_t \}_{t\geq 1}$ is a sequence of     $\mathrm{Uniform}(0,1)$ random variables independent of $\Hc_t$.
The vector $\btt(u) = \{\beta_0(u), \beta_1^*, \beta_2(u),\beta_3(u) - \beta_3(\tau_0) \}^\T$ satisfies that given the history $\Hc_t$, $q_t\{\btt(u) \}$ is strictly increasing with respect to $u \in (0,1)$.
For a fixed quantile level $\tau = \tau_0$, we have
\[
q_t(\btt^*) = \beta_0^* + \beta_1^*q_{t-1}(\btt^*) + \beta_2^*|Y_{t-1}|,~~e_t(\bta^*) = \theta_0^* + \theta_1^*e_{t-1}(\bta^*) + \theta_2^*|Y_{t-1}| + \theta_3^*|Y_{t-2}|,
\]
where $\bta^* = \{\int_0^{\tau_0} \beta_0(u)du, \tau_0\beta_1^*, \int_0^{\tau_0}\beta_2(u)du, \int_0^{\tau_0}\beta_3(u)du - \tau_0\beta_3(\tau_0)  \}^\T/\tau_0 $ and  $\btt^* = \{\beta_0(\tau_0), \beta_1^*, \beta_2(\tau_0) \}^\T$.
Here, $e_t(\bta^*)$ exhibits different dynamic behavior compared with $q_t(\btt^*)$ due to the extra lag quantity $|Y_{t-2}|$.
\end{remark}

\section{Methodology}\label{S3}
In this section, we propose a framework for estimating and performing statistical inference for the expected shortfall parameters $\bta^*$ in \eqref{var_es}.
We propose a two-step regression procedure for parameter estimation in Section \ref{OLS}.
 To perform model diagnostics, we construct a Portmanteau test to test whether the working conditional quantile and expected shortfall functions are correctly specified in Section \ref{LB}.

%an alternative two-step regression procedure in Section \ref{Sieve} based on a linear quantile approximation.
 %To further reduce the computational challenge for solving a non-linear quantile regression problem involved in Section \ref{OLS}, we propose 
 
%Section \ref{TWLS} introduces an approach to obtain sub-optimal weights when the conditional heteroscedasticity function is potentially misspecified.

\subsection{Weighted Two-Step Autoregressive ES Regression}\label{OLS}

%\kmcomment{I still do not like this.  I think you should just say you propose a two-step method for estimating the parameter of ES parameters in XXX.    Let S0 be xxxx (score function? is there a name that people use?) that depends on both beta and theta. Can be better written.  In general I think you shoudl avoid writing about the joint loss function here.  The joint loss will be roughly talked about in the introduction and then NOT appear anywhere else in the paper.   }

We propose to estimate the expected shortfall parameter $\bta^*$ using a two-step approach. 
Let $S_0(\btt, \bta; Y_t, \bX_t) = e_t(\bta) - q_t(\btt) - \mathbbm{1}\{Y_t \leq q_t(\btt)\}\{Y_t - q_t(\btt)\}/\tau$. 
This quantity can be interpreted as a score function for $\btt$ and $\bta$ under the joint quantile-expected shortfall framework in~\eqref{full_model}; see also \cite{H2023} and \cite{B2020} for similar score functions for the linear quantile-expected shortfall model.  
%within the framework of the linear quantile-expected shortfall model with $\mathrm{iid}$ data, as presented in \cite{H2023}. 
In the following proposition, we establish an orthogonality property for the score function $S_0(\btt, \bta; Y_t, \bX_t)$ in the context of an autoregressive quantile-expected shortfall model with time series data. % , the properties of $S_0(\btt, \bta; Y_t, \bX_t)$ are summarized in the following proposition.
\begin{proposition}\label{NO}
    For any time $t$, we have
    \[\EE\{S_0(\btt^*, \bta^*;Y_t, \bX_{t})\mid \Hc_t \}  =0\quad \mathrm{and}\quad  \frac{\partial \EE\{S_0(\btt, \bta;Y_t, \bX_{t})\mid\Hc_t \}}{\partial \btt}\bigg |_{\btt = \btt^*}=\bm{0}.\]
\end{proposition}
Proposition \ref{NO} states that,
given the history $\Hc_t$, $S_0(\btt, \bta; Y_t, \bX_{t})$ is an identification function for $\btt$ and $\bta$ 
and is locally robust to small perturbations of $\btt$ around the true value $\btt^*$.
%This motivates us to treat the quantile parameters $\btt$ as nuisance components.
Motivated by the above observation, the estimator for $\bta^*$ can be obtained by first estimating $\btt^*$ through solving a nonlinear quantile regression problem, followed by estimating $\bta^*$ using the score function $S_0(\btt, \bta; Y_t, \bX_t)$ with a plug-in quantile estimator $\hbt_T$.%\cite{B2020} and \cite{H2023}, and the above observation, 

%Let $G_1(\cdot)$ be an increasing and integrable function, $\mathcal{G}_2(\cdot)$ be a continuously three times differentiable with $G_2(\cdot) = \mathcal{G}_2'(\cdot)$, such that both $\mathcal{G}_2(\cdot)$ and $G_2(\cdot)$ are strictly positive. 
%Under \eqref{full_model}, the joint loss function is 
%\$
%S(\btt, \bta;Y_t, \bX_{t}) &= \left[\tau - \mathbbm{1}\left\{Y_t \leq q_t(\btt)\right\} \right] \left[G_1(Y_t) - G_1\left\{Y_t - q_t(\btt) \right\}\right]\\
%&+ G_2\{e_t(\bta) \} \left[ e_t(\bta) - q_t(\btt) - \frac{1}{\tau}\mathbbm{1}\{Y_t \leq q_t(\btt)\}\{Y_t - q_t(\btt)\}\right] - \mathcal{G}_2\left\{e_t(\bta) \right\},
%\$
%where $\bX_{t} = \{\bX_d \}_{d = 1}^{t-1}$.
%The estimators for $\btt^*$ and $\bta^*$ takes the form
%\begin{equation}\label{FZLoss}
%    \hbt_T,\hbta_T \in \argmin_{\btt, \bta} \frac{1}{T}\sum_{t = 1}^TS(\btt, \bta; Y_t, \bX_{t}).
%\end{equation}
%The asymptotic distributions of $\hbt_T$ and $\hbta_T$ are determined by the choices of $G_1(\cdot)$ and ${G}_2(\cdot)$. \cite{P2019} introduced a FZ0 loss function in \eqref{FZLoss} by picking $G_1(x) = 0$ and ${G}_2(x) = -1/x$ for $x<0$.
%They also considered similar dynamic structures of the conditional quantile and expected shortfall functions as in \eqref{var_es}.
%More specifically, 
Assume that the conditional quantile and expected shortfall functions $q_t(\btt)$ and $e_t(\bta)$ have autoregressive structures as in~\eqref{var_es}. 
In the first step, we propose to solve the following non-linear quantile regression problem:
 \begin{equation}\label{caviar}
 \hbt_T \in \underset{\btt \in \bm{B}}{\mathrm{argmin}}~\frac{1}{T}\sum_{t = 1}^T\rho_\tau\{Y_t - q_t({\btt})\},
 \end{equation}
 where $\rho_\tau(u) = u[\tau - \mathbbm{1}\{u \leq 0 \}]$ is the check loss function and $\bB$ is the parameter space for $\btt$.
 Given $\hbt_T$, we propose to estimate the expected shortfall parameters by minimizing the summation of the squared score function $S_0(\btt, \bta; Y_t, \bX_{t})$: % Given $\hbt_T$ obtained from \eqref{caviar}, the estimator $\hbta_T$ takes the form
 \begin{equation}\label{es_regression}
 \hbta_T \in \underset{\bta \in \bm{\Theta}}{\mathrm{argmin}}~\frac{1}{2T}\sum_{t = 1}^TS_0^2(\hbt_T, \bta;Y_t, \bX_{t}),
 \end{equation}
  where $\bTheta$ is the parameter space for $\bta$. 
  The problem~\eqref{es_regression} can be seen as a non-linear least squares problem with constructed pseudo-response variables $\{q_t(\hbt_T) + \mathbbm{1}\{Y_t \leq q_t(\hbt_T) \}\{Y_t - q_t(\hbt_T) \}/\tau \}_{t \geq 1}$.
  
The estimator $\hbta_T$ in~\eqref{es_regression} is generally not asymptotically efficient, i.e., it does not have the minimal asymptotic variance, due to the presence of conditional heteroskedasticity in the regression problem \eqref{es_regression}.
In the existing literature, a weighted least squares estimator has been shown to be generally more efficient than the unweighted least squares estimator, provided that the weighting function in the regression problem is set to be the inverse of the conditional variance of the regression error terms \citep{W1980,W2010}.
Hence, to achieve a more efficient estimator than $\hbta_T$, the key lies in consistently estimating the conditional heteroskedasticity terms in the regression problem \eqref{es_regression}.
It can be shown that
\begin{equation}\label{eq:conditional.variance}
\var\left\{S_0(\btt^*, \bta^*; Y_t, \bX_t) \mid \Hc_t\right\} = \frac{1}{\tau}\var\left\{ Y_t\mid Y_t\leq q_t(\btt^*), \Hc_t\right\} + \frac{1 - \tau}{\tau}\left\{q_t(\btt^*) - e_t(\bta^*)\right\}^2.
\end{equation}
However, there is no explicit parametric form of the term $\var\left\{ Y_t\mid Y_t\leq q_t(\btt^*), \Hc_t\right\}$ under the general data generating process given in \eqref{full_model}.
Estimating the conditional variance can be challenging without further modeling assumptions.

To this end, we propose a working model for the weighting function under the location-scale family time series.  Specifically, the time series $\{Y_t \}_{t\geq 1}$ can be written in the following form:
\begin{equation}\label{full_model_special}
    Y_t =  q_t(\btt^*) + \left\{q_t(\btt^*) - e_t(\bta^*) \right\}\zeta_t,
\end{equation}
where $Q_{\tau}(\zeta_t\mid \Hc_t)$ and $\ES_{\tau}(\zeta_t\mid \Hc_t)$ are $0$ and $-1$, respectively.
As shown in Proposition S5.1 of Section S5.1 in the online supplementary material, \eqref{full_model_special} is equivalent to a general class of location-scale time series models with independent and identically distributed (\textit{i.i.d.}) innovations, provided that $\{\zeta_t\}_{t \geq 1}$ is an \textit{i.i.d.}~sequence of appropriately scaled random variables and $\zeta_t$ is independent of $\Hc_t$ for any $t$.
Under the \textit{i.i.d.}~assumption on the innovation process, the conditional variance of the regression error terms in \eqref{es_regression} can be expressed as
\begin{equation}\label{hetero}
\var\{S_0(\btt^*, \bta^*;Y_t, \bX_{t}) \mid \Hc_t\} = \{q_t(\btt^*) - e_t(\bta^*)\}^2\left\{\frac{1}{\tau}\EE(\zeta_t^2\mid \zeta_t\leq 0) - 1 \right\}.
\end{equation}
Hence, $\var\{S_0(\btt^*, \bta^*;Y_t, \bX_{t}) \mid \Hc_t\} \propto \{q_t(\btt^*) - e_t(\bta^*) \}^2$ for $t \geq 1$.
Motivated by the above derivation,
we set the weighting function for the $t$-th observation as $1/\{q_t(\btt^*) - e_t(\bta^*) \}^2$ with the corresponding true parameters $ \{(\btt^*)^\T,(\bta^*)^\T\}^\T$.

Let $\hbt_T$ and $\hbta_T$ be estimators obtained from solving \eqref{caviar} and \eqref{es_regression}, respectively.
We propose a weighted expected shortfall estimator $\hbta^W_{T}$ that takes the following form:
 \begin{equation}\label{es_regression_wls}
 \hbta_T^W \in \argmin_{\bta \in \bm{\Theta}}~\frac{1}{2T}\sum_{t = 1}^T{\frac{1}{\{q_t(\hbt_T) - e_t(\hbta_T) \}^2 }}S_0^2(\hbt_T, \bta;Y_t, \bX_t).
 \end{equation} 
%Given $\hbta_T^W$, 
For any $j \in \{ 1,2,\ldots, d_e\}$ and $\alpha \in (0,1)$, the $1-\alpha$-level confidence interval for $\theta_j^*$ can then be constructed as
\begin{equation}\label{CI_wls}\left[\hat{\theta}_{T,j}^W\pm\Phi^{-1}\left(1-\frac{\alpha}{2}\right)\sqrt{\frac{1}{T}\left\{(\hat{\bSigma}_T^W)^{-1}\hat{\bOmega}_T^W(\hat{\bSigma}_T^W)^{-1}\right\}_{jj}}\right],\end{equation}
where $\hat{\bOmega}_T^W = \sum_{t = 1}^TS_0^2(\hbt_T, \hbta_T^W;Y_t, \bX_{t})\nabla e_t(\hbta_T^W)\nabla e_t (\hbta_T^W)^{\T}/[T \{q_t(\hbt_T) - e_t(\hbta_T) \}^4]$ and $\hat{\bSigma}_T^W = \sum_{t = 1}^T \nabla e_t(\hbta_T^W)\nabla e_t(\hbta_T^W)^{\T}/[T{\{q_t(\hbt_T) - e_t(\hbta_T) \}^2}]$, and $\Phi(\cdot)$ is the cumulative distribution function for a standard normal distribution.

    Additionally, we also consider an alternative approach to construct confidence intervals based on random weighting bootstrap in Section~\ref{S5} for numerical stability.
    We defer the details of this construction to Section S2.2 of the online supplementary material.
Moreover, it should be noted that while our choice of the weighting function $1/\{q_t(\btt^*) - e_t(\bta^*) \}^2$ is motivated by the location-scale time series models with \textit{i.i.d.}~innovations, the validity of the proposed inference procedure does not rely on such specific data-generating assumptions, as illustrated by Theorems \ref{normality_wls}--\ref{VCW} of Section~\ref{S4}.
Under such an \textit{i.i.d.}~assumption, our weighted estimator attains asymptotic efficiency within the class of joint $M$-estimators considered in \cite{FZ2016}, as demonstrated in Corollary \ref{ARE} of Section~\ref{S4}.

Finally, we provide a generic algorithm to solve \eqref{es_regression_wls}.
Our optimization steps include Broyden–Fletcher–Goldfarb–Shanno (BFGS) \citep{F1964} and Nelder-Mead \citep{N1965} algorithms.
Our implementation is based on applying \texttt{optim} function in \texttt{R}.
We note that \eqref{caviar} and \eqref{es_regression} can be solved by slight modifications of Algorithm~\ref{Alg0}.

\begin{remark}
    The optimal weighting function derived in this section relies on the assumption that the data are generated from a location-scale family time series model with \textit{i.i.d.} innovations. The weighting function is no longer optimal when this assumption is violated. However, since the term $\{q_t(\btt^*)  - e_t(\bta^*) \}^2$ remains an important component of~\eqref{eq:conditional.variance}, our weighting function can still be viewed as a working approximation to the optimal weights when~\eqref{full_model_special} no longer holds. We conduct numerical experiments to quantify the efficiency loss of our weighted estimator~\eqref{es_regression_wls} when the data-generating process deviates from the location-scale family with \textit{i.i.d.} innovations. Our results show that, even when the working weighting function is suboptimal in such settings, the efficiency loss is very small. See Section 5, Model~\eqref{eq:global.model}, and Section S9 of the online supplementary material for details.
\end{remark}

\begin{algorithm}[!ht]
\caption{A Framework for Solving \eqref{es_regression_wls}}
\label{Alg0}
\begin{algorithmic}[0]

\STATE \textbf{Step 1 :}
\STATE \hspace{1em}Randomly sample $10{,}000$ initial parameter vectors from a uniform distribution.

\STATE \textbf{Step 2 :}
\STATE \hspace{1em}Among the $10{,}000$ initial parameters, select $10$ vectors $\{\hbta_{(1)}^0, \ldots ,\hbta_{(10)}^0 \}$ that return the lowest empirical loss value in \eqref{es_regression_wls}.

\STATE \textbf{Step 3 :}
\FOR{$i = 1,\ldots,10$}

    \STATE \textit{Initialize:} Set the starting value $\hat{\bta}_{(i)}' \leftarrow\hat{\boldsymbol{\theta}}_{(i)}^{0}$.

    \STATE \textit{Refinement loop:} Repeat the following procedure $5$ times and output $\hbta_{(i)}$:
    \begin{enumerate}[itemsep=0pt,parsep=0pt,topsep=0pt,label=(\roman*)]
        \item \textbf{BFGS update:}  Obtain $\hbta_{(i)}$ by optimizing \eqref{es_regression_wls} using the BFGS algorithm initialized at $\hat{\boldsymbol{\theta}}_{(i)}'$.
        \item \textbf{Perturbation:} Add a small Gaussian perturbation
        \[
        \hat{\boldsymbol{\theta}}_{(i)}' \leftarrow \hat{\boldsymbol{\theta}}_{(i)} + \boldsymbol{\iota},
        \qquad
        \boldsymbol{\iota} \sim \mathcal{N}(\boldsymbol{0},\,0.1\mathbf{I}).
        \]        
        \item \textbf{Nelder--Mead update:} Obtain $\hbta_{(i)}$ by optimizing \eqref{es_regression_wls} with Nelder-Mead algorithm initialized at $\hat{\boldsymbol{\theta}}_{(i)}'$.
        \item \textbf{Perturbation:} Repeat step (ii).
    \end{enumerate}

\ENDFOR

\STATE \textbf{Step 4 :}
\STATE \hspace{1em}Compute the empirical loss in \eqref{es_regression_wls} for
$\{\hat{\boldsymbol{\theta}}_{(1)}, \ldots, \hat{\boldsymbol{\theta}}_{(10)}\}$. Select the final estimator $\hat{\boldsymbol{\theta}}_{T}^W$ as the vector achieving the smallest empirical loss.

\end{algorithmic}
\end{algorithm}

%\textcolor{red}{\textit{i.i.d.}~needs to be followed by a tilde.} 

\subsection{Model Diagnostics via a Portmanteau Test}\label{LB}
Under the autoregressive quantile-ES framework in~\eqref{var_es}, it is important to accurately determine the appropriate autoregressive orders and functional forms of $g(\cdot)$ and $h(\cdot)$.
In this section, we assume that the data $\{Y_t \}_{t \geq 1}$ are generated by the location-scale family time series with \textit{i.i.d.}~innovations.
Under this assumption,
we develop a Portmanteau test for model diagnostics  \citep{L1994,L2005}.
%In practice, assessing the validity of the proposed working model is of great interest. Under the general autoregressive quantile-expected shortfall framework \eqref{var_es}, it is important to accurately determine the appropriate autoregressive orders and the functional forms of $g(\cdot)$ and $h(\cdot)$.
%A valid statistical test is needed for us to make the correct model specification.

%In this section, we consider the location-scale time series model \eqref{full_model_special} and introduce a Portmanteau test motivated by \cite{L1994} and \cite{L2005}.
Let $\zeta_t = \{Y_t - q_t(\btt^*)\}/\{q_t(\btt^*) - e_t(\bta^*) \}$.
Under the assumption that $\{Y_t\}_{t \geq 1}$ is generated by the location-scale family time series with \textit{i.i.d.}~innovations, and that both quantile and expected shortfall functions are correctly specified with true parameters $\btt^*$ and $\bta^*$ known, it can be shown that $\{ \zeta_t\}_{t \geq 1}$ is a sequence of \textit{i.i.d.}~random variables, see details in Proposition S5.1 in the online supplementary material.
Therefore, the dependence structure of $\{\zeta_t \}_{t \geq 1}$ reflects whether $q_t(\btt^*)$ and $e_t(\bta^*)$ correctly specify the conditional quantile and expected shortfall of $Y_t$, respectively. Motivated by this observation, we propose the following model diagnostic procedure based on detecting serial dependence in $\{\zeta_t \}_{t\geq 1}$.

For any positive integer $m$, let $r_m = \mathrm{corr}(\zeta_t^2, \zeta_{t-m}^2)$.  We develop a Portmanteau test to test the following hypothesis problem:
\begin{equation}\label{Port_H0_raw}
    \mathrm{H}_0: r_m = 0~\forall~m\geq 1~~\mathrm{versus}~~\mathrm{H_a}: \exists~m \geq 1~\mathrm{such~that~} r_m \neq 0.
\end{equation}
In practice, it is infeasible to examine all lag levels. 
To this end, we pick a positive integer $M$ and consider testing the hypothesis problem
\begin{equation}\label{Port_H0_M}
    \mathrm{H}_0: r_m = 0,~\forall~m \in [M]~~\mathrm{versus}~~\mathrm{H_a}: \exists~m \in [M]~\mathrm{such~that~} r_m \neq 0,
\end{equation}
where $[M] = \{1,\ldots,M \}$.
The integer $M$ should be chosen such that $Y_t$ and $Y_{t+M}$ are approximately independent. Common choices of $M$ include $\{4, 6,8,10 \}$. Our simulation results in Section 5 show that the empirical performance of the Portmanteau test is robust to the choice of $M$. In the real data applications in Section~6, we set $M = 10$, which corresponds to the length of two trading weeks.
We now outline the procedure for performing hypothesis testing on~\eqref{Port_H0_M}; detailed derivations of the proposed testing procedure are deferred to Section S2.3 of the online supplementary material.
The test is based on the estimators $\hbt_T$ and $\hbta_T$ obtained by \eqref{caviar} and \eqref{es_regression}, respectively. 
We construct the estimated noise term as
$
\hat{\zeta}_t ={Y_t - q_t(\hbt_T)}/\{q_t(\hbt_T) - e_t(\hbta_T)\}.
$

Let $\tilde{\br} = (\tilde{r}_1, \ldots, \tilde{r}_M )^\T$ be the vector of the sample autocorrelation coefficients of $\{\hat{\zeta}_t^2 \}_{t \geq 1}$ up to the lag $M$.
The derivation steps given in the proof of Theorem \ref{port_theory} show that under $\mathrm{H}_0$, $\sqrt{T}\tilde{\br}$ is asymptotically normally distributed with mean $\bm 0$ and variance $\bOmega^{\mathrm{test}}$.
Furthermore, $T\tilde{\br}^\T(\bOmega^{\mathrm{test}})^{-1}\tilde{\br}$
converges in distribution to a $\chi_M^2$ random variable.
The general procedure for testing \eqref{Port_H0_M} is summarized in Algorithm \ref{AlgBT_simp}.
The closed form expression of $\bOmega^{\mathrm{test}}$ involves many intermediate quantities and is omitted from the main manuscript. We refer the reader to Section S2.3 in the online supplementary material for the details on the derivation and estimation of $\bOmega^{\mathrm{test}}$. 
\begin{algorithm}[!ht]
    \caption{The Portmanteau Test for Testing \eqref{Port_H0_M}.} 
    \begin{algorithmic}[0] \label{AlgBT_simp}
	\STATE \textbf{Step1 :} Set a positive integer $M$. Typical choices of $M$ include $\{4, 6,8,10 \}$.
 \STATE \textbf{Step2 :} Obtain $\hbt_T$ and $\hbta_T$ by solving \eqref{caviar} and \eqref{es_regression} with dataset $\{Y_t, \bX_t \}_{t \geq  1}$.
 \STATE \textbf{Step3 :} Compute $\{\hat{\zeta}_t\}_{t \geq  1} =  [\{ Y_t - q_t(\hbt_T)\}/\{q_t(\hbt_T) - e_t(\hbta_T) \}]_{t \geq 1}$ and $\tilde{\br}$.
    \STATE \textbf{Step4 :} Estimate $\hat{\bOmega}^{\mathrm{test}}$ from data.
    \STATE \textbf{Step5 :} Let $\hat{\psi} = T\tilde{\br}^\T(\hat{\bOmega}^{\mathrm{test}})^{-1} \tilde{\br}. $ Reject the null hypothesis in \eqref{Port_H0_M} if the test statistic $\hat{\psi}\geq \chi^2_{M, 1 - \alpha}$ for a given confidence level $\alpha \in (0,1)$.
	    \end{algorithmic}
    \end{algorithm}

\section{Theoretical Results}\label{S4}
We study the theoretical properties of the proposed two-step expected shortfall regression estimators in Section \ref{S3}.
We first show that the unweighted expected shortfall estimator $\hbta_T$ is weakly consistent, and provide the convergence rate and asymptotic normality of $\hbta_T$ in Section \ref{S4.1}.
Parallel theoretical results for the weighted estimator $\hbta_T^W$ are presented in Section \ref{S4.2}.
The theoretical guarantee of the proposed Portmanteau test is included in Section~\ref{S4.3}.
 We defer all assumptions required for the theoretical results to Section S3 of the online supplementary material.

We start with defining some notation.
For a vector $\br = (r_1, \ldots, r_p)^\T \in \mathbb{R}^p$, let $ \lVert \br \rVert_q = (\sum_{j = 1}^p |r_j|^q )^{1/q}$ for $1\leq q < \infty$.
For sequences $\{a_n \}$ and $\{b_n\}$, we write $a_n = O(b_n)$ if there exists a constant $C$ such that $a_n \leq Cb_n$ for all $n$ and $a_n = O_p(b_n)$ if $a_n/b_n = O_p(1)$.
We write $a_n = o(b_n)$ if $\lim_{n\rightarrow\infty}a_n/b_n = 0$ and $a_n = o_p(b_n)$ if $a_n/b_n \stackrel{p}{\rightarrow} 0$.
Let $\EE_t(\cdot) = \EE(\cdot \mid\Hc_t)$.
For a function $f(\btt): \mathbb{R}^{\mathrm{dim}(\btt)}\rightarrow \mathbb{R}$, let $\nabla f(\btt) = \partial f(\btt)/\partial \btt$ be the gradient of $f(\btt)$ with respect to the parameter $\btt$.

%We start with defining some notation.
%For a vector $\br = (r_1, \ldots, r_p)^\T \in \mathbb{R}^p$, let $ \lVert \br \rVert_q = (\sum_{j = 1}^p r_j^q )^{1/q}$ for $1\leq q < \infty$. For a $m$ by $n$ matrix ${\Ab}$, let $\lVert {\Ab} \rVert_F = \sqrt{\sum_{1\leq j \leq m,1\leq k \leq n}A_{jk}^2}$ be the Frobenius norm of ${\Ab}$ and $\lVert {\Ab}\rVert_{\max} = \max_{1\leq j \leq m,1\leq k \leq n}| A_{jk}|$ to be the maximum absolute entry value of ${\Ab}$, let $\lambda_{\min}(\Ab)$ be the minimal eigenvalue of $\Ab$.
%For sequences $\{a_n \}$ and $\{b_n\}$, we write $a_n = O(b_n)$ if there exists a constant $C$ such that $a_n \leq Cb_n$ for all $n$ and $a_n = O_p(b_n)$ if $a_n/b_n = O_p(1)$.
%Let $a_n = o(b_n)$ if $\lim_{n\rightarrow\infty}a_n/b_n = 0$ and $a_n = o_p(b_n)$ if $a_n/b_n \stackrel{p}{\rightarrow} 0$.
%For a positive integer $N$, let $[N]$ be the set $\{1, 2, \ldots, N \}$.
%Let $\EE_t(\cdot) = \EE(\cdot \mid\Hc_t)$.
%For a function $f(\btt): \mathbb{R}^{\mathrm{dim}(\btt)}\rightarrow \mathbb{R}$, let $\nabla f(\btt) = \partial f(\btt)/\partial \btt$ be the gradient of $f(\btt)$ with respect to the parameter $\btt$.
\subsection{ Unweighted Expected Shortfall Estimator}\label{S4.1}
We derive the asymptotic properties of $\hbta_T$ obtained by solving \eqref{es_regression}. 
The first step of our proposed approach involves solving an autoregressive quantile regression problem \eqref{caviar}.
Let 
% which can be expressed by the event:
\begin{equation}\label{beta_consistency}
    \mathcal{E}_{1T} = \left\{\hbt_T \in \mathbb{B}(\btt^*,r_T) = \left\{ \btt \in \bm{B};~\norm{\btt - \btt^*}_2\leq r_T\right\};~r_T = o(T^{-\frac{1}{4}}) \right\}
\end{equation}
be an event characterizing the convergence rate of quantile estimator $\hbt_T$.  
The event $\mathcal{E}_{1T}$ has been shown to hold with high probability tending to $1$ as $T$ increases \citep{X2021}, which implies that $\hbt_T$ is weakly consistent with a mild rate.
In the following, we will conduct our analysis, conditioned on the high-probability event $\mathcal{E}_{1T}$. 

We start by showing that $\hbta_T$ obtained from \eqref{es_regression} is weakly consistent. 
Our theoretical results require a series of regularity assumptions on the smoothness of quantile-expected shortfall functions, the existence of certain envelope functions, the stationarity and ergodicity conditions, and identifiability conditions.
Specifically, our theoretical analysis requires the existence of sixth-moment bounds on $Y_t$ and some corresponding envelope functions. This assumption allows the data to be heavy-tailed to a certain extent, but not without limitation. We refer readers to Assumption S1 of the online supplementary material for details.
%\textcolor{red}{Section XXX of the online supplementary material. Probably name it Assumption S1 so that it is clear that it is in the supp?}. 

%theoretical result builds upon Assumption \ref{sp-ESC}, which imposes mild regularity conditions on the regression problem \eqref{es_regression}. These conditions include the smoothness of quantile-expected shortfall functions, the existence of certain envelope functions and identifiability conditions. 
%The details regarding Assumption \ref{sp-ESC} are provided in Appendix S3.

 \begin{theorem}\label{consistency}
Under Assumption S1 and the event $\mathcal{E}_{1T}$, the estimator $\hbta_T$ is weakly consistent: $
 \hbta_T \stackrel{p}{\longrightarrow} \bta^*.
 $
 \end{theorem}

%The proof of Theorem \ref{consistency} is standard and follows a similar manner in \cite{W1996} and \cite{N1994}.

%Assumption \ref{sp-ESC} provides some mild regularity conditions to ensure the consistency result of the estimator $\hbta$, it combines some similar arguments made in Assumption C in \cite{E2004} and Condition M in \cite{B2020}. Assumptions (1.1) and (1.2) are standard assumptions on sample space of $Y_t$ and smoothness on ES function $e_t(\bta)$. Assumption (1.3) is used to ensure the sample loss function $1/T \sum_{t = 1}^TS_0^2(\btt, \bta, Y_t, \bX_t)$ will converge to the population loss function by probability. This assumption have been extensively applied in non linear least square regression related papers (\citet{A1983,N1994, W1996}). In the context of time series and dependent data, this assumption can be satisfied by various forms of uniform law of large numbers, see examples in \cite{A1987} and \cite{P1989}. Assumption (1.4) introduces some "envelope functions" to control the second moment bounds for the quantile function $q_t(\btt)$ and ES function $e_t(\bta)$, which serves as sufficient conditions to make sure the population loss function $\EE\{1/T \sum_{t = 1}^TS_0^2(\btt, \bta, Y_t, \bX_t) \}$ is well defined, which also mitigates potential explosive tendencies of $Y_t$. Assumption (1.5) imposed the identifiability condition for $\bta^*$, which is analogous to Assumption C7 in \cite{E2004}.

Next, we characterize an upper bound for the estimation error of $\hbta_T$ under the $\ell_2$ norm and establish its asymptotic distribution.
For notational simplicity, let $S_t(\btt, \bta) = S_0(\btt, \bta; Y_t, \bX_t)$.
We define the event $\mathcal{E}_{2T}$ as:
%As discussed in Section \ref{S3}, we constructed a sequence of pseudo-response variables to solve the regression problem \eqref{es_regression}. 
%To ensure the validity of this construction, the following event is imposed.
\begin{equation*}
\left\{\sup_{\btt \in \mathbb{B}(\btt^*,r_T)}\norm{\frac{1}{T} \sum_{t = 1}^T(1 - \EE_t)\left\{S_t(\btt, \bta^*)\nabla e_t(\bta^*) - S_t(\btt^*, \bta^*)\nabla e_t(\bta^*)\right\}}_2 \leq O\left(\frac{s_T}{\sqrt{T}}\right)\right\},
\end{equation*}
where $\lim_{T \rightarrow \infty}s_T = 0$.
The event $\mathcal{E}_{2T}$ establishes the stochastic equicontinuity condition for the quantity ${1}/{T} \sum_{t=1}^T (1 - \EE_t){S_0(\btt, \bta^*; Y_t, \bX_t)\nabla e_t(\bta^*)}$. 
This property is a crucial condition to obtain the estimation error bound and the asymptotic normality of $\hbta_T$. 
This condition can be verified to hold under some sufficient conditions. 
In Section S4 of the online supplementary material, we provide a proof demonstrating that $\PP(\mathcal{E}_{2T}) \rightarrow 1$ under certain moment bounds and convergence rate conditions.
The derivation of the asymptotic properties of $\hbta_T$ relies on an upper bound for the conditional density of the quantile error $\epsilon_t$ and the existence of a positive definite covariance matrix.
These requirements are summarized in Assumption S2.

 \begin{theorem}\label{normality}
     Under Assumptions S1--S2 and the event $\mathcal{E}_{1T}\cap \mathcal{E}_{2T}$, we have
         (i) $\norm{\hbta_T - \bta^*}_2 = O_p\left\{{(1 + s_T) }/{\sqrt{T}}  + r_T^2\right\}$ and
         (ii) $\sqrt{T} \bOmega_T^{-\frac{1}{2}}\bSigma_{T}\left(\hbta_T - \bta^*\right)\stackrel{d}{\longrightarrow} \mathcal{N}\left(\bm{0}, \bI_{d_e} \right).$        
 \end{theorem}

The conclusion of Theorem \ref{normality} consists of two parts.
The first part characterizes the convergence rate of $\hbta_T$. Our results demonstrate that the estimation error induced by $\hbt_T$ is first-order negligible and does not affect the asymptotic distribution of $\hbta_T$. This finding aligns with the linear expected shortfall regression results in \cite{H2023}.
The second part establishes the asymptotic normality of $\hbta_T$. The variance-covariance matrix $\bSigma_T^{-1}\bOmega_T\bSigma_T^{-1}$ exhibits a structure similar to that of the results in \cite{P2019}, although they are not asymptotically equivalent.

\subsection{ Weighted Expected Shortfall Estimator}\label{S4.2}
In this section, we derive the asymptotic properties of the weighted estimator $\hbta_T^W$ obtained by solving \eqref{es_regression_wls}.
%We define a $\Hc_t$-measurable weighting function as $ 1/\{q_t(\btt^*) - e_t(\bta^*) \}^2$ for $S_0^2(\hbt_T, \bta; Y_t, \bX_t)$. %. , following Section \ref{S3}.
%The true parameter vector  $ \{(\btt^*)^\T, (\bta^*)^\T \}^\T$ can be estimated from the data as $(\hbt_T^\T, \hbta_T^\T)^\T$.
We start by imposing two high-level events:
\begin{equation}\label{var_consistency}
    \mathcal{E}_{3T} = \left\{\left(\hbt_T^\T, \hbta_T^\T\right)^\T \in \mathbb{B}[\{(\btt^*)^\T, (\bta^*)^\T \}^\T ,d_T];~d_T = o(1) \right\}
\end{equation}
and the event $\mathcal{E}_{4T}$ is defined as
\begin{equation*}
    \left\{\sup_{(\btt^\T ,\bta^\T)^\T \in \BB^*}\norm{\frac{1}{T} \sum_{t = 1}^T(1 - \EE_t)\left\{\frac{S_t(\btt, \bta^*)\nabla e_t(\bta^*)}{\{q_t(\btt) - e_t(\bta) \}^2} - \frac{S_t(\btt^*, \bta^*)\nabla e_t(\bta^*)}{\{ q_t(\btt^*) - e_t(\bta^*)\}^2}\right\}}_2 \leq O\left(\frac{s_T}{\sqrt{T}}\right)\right\}.
\end{equation*}
Here, we use $\BB^*$ to denote $\mathbb{B}[\{(\btt^*)^\T, (\bta^*)^\T \}^\T,d_T]$ for notational simplicity.
When the parameters for the weighting function can be consistently estimated, the event $\mathcal{E}_{3T}$ holds with high probability converging to one as $T$ increases, which is a common assumption required by weighted least squares in the existing literature \citep{W1980,R2017}.
For a given model specification \eqref{full_model_special}, $\mathcal{E}_{3T}$ can be satisfied through $\mathcal{E}_{1T}$ and Theorem \ref{consistency}.
The event $\mathcal{E}_{4T}$ extends $\mathcal{E}_{2T}$ by incorporating the weighting functions. 
A detailed discussion of the validity of $\mathcal{E}_{4T}$ is provided in Section S4 of the online supplementary material.

%The event $\mathcal{E}_{4T}$ is generalizes $\mathcal{E}_{2T}$ with weighting function $w_t(\bgamma)$.

In the following, we derive a weak consistency result for the weighted least squares expected shortfall estimator $\hbta^W_T$.
%the weak consistency result of $\hbta_T$ that is similar in Theorem \ref{ESC}.
The regularity conditions are in Assumption S3, which generalizes Assumption S1 by incorporating the weighting functions.

\begin{theorem}\label{wls_consistency}
Under Assumptions S1--S3 and the event $\mathcal{E}_{1T}\cap \mathcal{E}_{3T}$, the estimator $\hbta^W_T$ is weakly consistent:
$
\hbta^W_T \stackrel{p}{\longrightarrow} \bta^*.
$    
\end{theorem}

In Section \ref{OLS}, we estimate $\bta$ using an unweighted least squares regression, where $\btt^*$ is replaced by $\hbt_T$ in the score function $S_0(\btt, \bta; Y_t, \bX_t)$. The orthogonality property for $\btt$ is established in Proposition \ref{NO}.
We extend this orthogonality condition for the weighted least squares regression problem \eqref{es_regression_wls}, as formalized in Proposition S5.2 in Section S5.1 of the online supplementary material.
Based on Proposition S5.2, we analyze the asymptotic behaviors of $\hbta_T^W$. 
The derivation follows closely from the proof of Theorem \ref{normality}.
%To establish asymptotic normality for $\hbta_T^W$,  we impose regularity conditions as detailed in Assumption S3. 
%This assumption generalizes Assumption S2 by including the weighting functions. 
%Consequently, additional moment bounds and weak convergence conditions are required.

\begin{theorem}\label{normality_wls}
    Under Assumptions S1--S3 and the event $\mathcal{E}_{1T}\cap \mathcal{E}_{3T}\cap\mathcal{E}_{4T}$, we have
         (i) $\norm{\hbta_T^W - \bta^*}_2 = O_p\left\{{(1 + s_T)}/{\sqrt{T}}  + r_T^2\right\}$; and
         (ii) $\sqrt{T}\left(\bOmega_T^W\right)^{-\frac{1}{2}}\bSigma_{T}^W \left(\hbta_T^W - \bta^*\right)\stackrel{d}{\longrightarrow} \mathcal{N}\left(\bm{0}, \bI_{d_e} \right).$
\end{theorem}

Theorem \ref{normality_wls} shows that the estimation errors of the components of the quantile and weighting function do not affect the asymptotic distribution of $\hbta_T^W$.
The second part indicates that the variance-covariance matrix of $\hbta_T^W$ differs from that of Theorem \ref{normality} due to the existence of the weighting functions.
A properly constructed weighting function can improve the asymptotic efficiency of $\hbta_T^W$, leading to narrower confidence intervals for statistical inference in $\bta^*$. This result is summarized in the following corollary.

\begin{corollary}\label{ARE}
Suppose that $\{q_t(\btt^*)  - e_t(\bta^*)\}^2 \propto \var\{S_0(\btt^*, \bta^*;Y_t, \bX_{t})\mid\Hc_t\}$ and Theorem \ref{normality_wls} hold. Then, the estimator $\hat{\bta}_T^W$ is asymptotically efficient among the class of joint $M$-estimators introduced in \cite{FZ2016}.
\end{corollary}

In the following theorem, we show that our proposed variance-covariance matrix estimator given in Section \ref{OLS} is consistent.
%To this end, we impose additional moment bound restrictions, which we detail in Assumption \ref{sp-VCWasump}.

\begin{theorem}\label{VCW}
Under Assumptions S1--S3 and the event $\mathcal{E}_{1T} \cap \mathcal{E}_{2T}\cap \mathcal{E}_{3T}$, the estimator of the variance-covariance matrix in Theorem \ref{normality_wls} is weakly consistent:
\begin{equation*}
\left(\hat{\bSigma}_T^W \right)^{-1}\hat{\bOmega}_T^W\left(\hat{\bSigma}_T^W \right)^{-1} \stackrel{p}{\longrightarrow} \left(\bSigma_T^W\right)^{-1}\bOmega_T^W\left(\bSigma_T^W\right)^{-1}.
\end{equation*}
\end{theorem}
The above theorem validates the inference procedure given in \eqref{CI_wls}.
Note that Theorem \ref{VCW} also provides theoretical guarantees for the confidence intervals given in Theorem \ref{normality} by setting the weighting function equal to one for every $t$, hence the variance-covariance matrix of $\hbta_T^W$ will reduce to that of $\hbta_T$.

\subsection{Validity of Portmanteau Test}\label{S4.3}
In this section, we provide theoretical guarantees for the Portmanteau test proposed in Section \ref{LB} for model diagnostics. Specifically, we characterize the asymptotic behavior of the test statistic $\hat{\psi}$ introduced in Algorithm \ref{AlgBT_simp}, under the null hypothesis $\mathrm{H}_0$ in \eqref{Port_H0_M}.
The following two theorems establish the validity of the proposed Portmanteau test. 
%To this end, we impose conditions on moment bounds and convergence rates for certain intermediate quantities in Assumptions \ref{sp-Port_asmp}--\ref{sp-asump_omegaconsistent}, detailed in Section S3 in the online supplementary material.  
%indicate the validity of our proposed Portmanteau test.

\begin{theorem}\label{port_theory}
    Suppose $\{Y_t\}_{t\geq 1}$ is generated by the location-scale family model \eqref{full_model_special} and let $\{\hat{\zeta}_t \}_{t \geq 1} = [\{Y_t - q_t(\hbt_T) \}/\{ q_t(\hbt_T) - e_t(\hbta_T)\}]_{t \geq 1}$ where $\hbt_T$ and $\hbta_T$ are obtained by solving \eqref{caviar} and \eqref{es_regression}, respectively. For a fixed constant $M$, let $\tilde{\br} =(\tilde{r}_1, \tilde{r}_2, \ldots, \tilde{r}_M)^\T$ be the vector of sample autocorrelation coefficients of $\{ \hat{\zeta}_t^2\}_{t \geq1}$ up to lag level $M$.
    Then for the matrix $\bOmega^{\mathrm{test}}$ defined in (S5.13) in the online supplementary material, under Assumptions~S1--S3, and the event $\mathcal{E}_{1T}\cap \mathcal{E}_{2T}$, we have $T\tilde{\br}^\T\left(\bOmega^{\mathrm{test}} \right)^{-1}\tilde{\br} \stackrel{d}{\longrightarrow} \chi^2_M.
    $
\end{theorem}

\begin{theorem}\label{omegahatconsistent}
   Let $\hat{\bOmega}^{\mathrm{test}}$ be the estimator given in Algorithm S2 in the online supplementary material.
   Under Assumptions S1--S3 and the event $\mathcal{E}_{1T}\cap \mathcal{E}_{2T}$, we have $
   \hat{\bOmega}^{\mathrm{test}} \stackrel{p}{\longrightarrow}\bOmega^{\mathrm{test}}.
   $
\end{theorem}

\section{Numerical Studies}\label{S5}
We conduct extensive numerical experiments to assess the finite-sample performance of our proposed expected shortfall estimator $\hbta_T^W$ obtained from \eqref{es_regression_wls}.
We also compare to a class of estimators obtained from minimizing the different joint loss functions given in \cite{P2019} and \cite{DF2020}. 
Section \ref{S5.1} introduces the data generation processes.
%Section  describes implementation details of the experiments.
Section \ref{S5.3} presents the results for the different estimators and illustrates the efficiency gain of the proposed weighted expected shortfall estimator relative to existing approaches.
%Section \ref{sec:time.consuming} evaluates the computational efficiency of our proposed estimators.
We conduct numerical studies to assess the performance of the proposed Portmanteau test for model diagnostics in Section \ref{S5.5}. % conducts model diagnostics by applying the testing procedure given in Section \ref{LB}.
%Section \ref{S5.4} compares the time consumption for our two-step approach and the approach in \cite{P2019}.

\subsection{Data Generating Process and Methodology Comparison}\label{S5.1}
We start with two data-generating processes for numerical studies: %considered in Section \ref{S5.3}:
%(i) a linear generalized autoregressive conditional heteroskedastic model;
(i) an asymmetric slope model; and
(ii) a non-location-scale family symmetric absolute value model.
The above models correspond to scenarios in which the weighting function in \eqref{es_regression_wls} is optimal and suboptimal, respectively.
The specifications of the two models are:
\begin{enumerate}
    %\item[(i)] The linear generalized autoregressive conditional heteroskedastic model is widely considered in many financial applications \citep{X2009}.
    %Let
    %\begin{align}\label{GARCH_simu}
    %Y_t &= \sigma_t(\balpha^*)\eta_t,~~\sigma_t(\balpha^*) = \alpha_0^* + \alpha_1^* \sigma_{t-1}(\balpha^*) + \alpha_2^*|Y_{t-1}|,\\ \notag
    %& q_t(\btt^*) = \beta_0^* + \beta_1^*q_{t-1}(\btt^*) + \beta_2^*|Y_{t-1}|,~~e_t(\bta^*) = \theta_0^* + \theta_1^*e_{t-1}(\bta^*) + \theta_2^*|Y_{t-1}|,\notag
%\end{align}
 %where $\{\eta_t \}_{t \geq 1}$ are  \textit{i.i.d.}~standard normal.
 %The parameters can be shown to take the forms  $\btt^* = \{\alpha_1^*Q_\tau(\eta), \alpha_2^*, \alpha_3^*Q_\tau(\eta) \}^\T$ and $\bta^* = \{\alpha_1^*\ES_{\tau}(\eta),\alpha_2^*, \alpha_3^*\ES_\tau(\eta) \}^\T$, respectively.
 %Note that under \eqref{GARCH_simu}, $q_t(\btt^*)$ and $e_t(\bta^*)$ follow the symmetric absolute value model in Example \ref{SAVExp} in Section \ref{S2}.
    \item[(i)]  The first data generating process combines the quasi-score driven model in \cite{B2023} and the asymmetric slope model given in \cite{E2004}. Recall from Section \ref{S2} that $Y_t^+ = Y_t \mathbbm{1}\{Y_t \geq 0 \}$ and $Y_t^- = Y_t\mathbbm{1}\{Y_t < 0 \}$.
     Let $f_t(\balpha^*) = \alpha_0^* + \alpha_1^*f_{t-1}(\balpha^*) + \alpha_2^*Y_{t-1}^+ - \alpha_3^*Y_{t-1}^-$ and
    \begin{equation}\label{AS_simu}
    Y_t = f_t(\balpha^*) + (\alpha_4^* + \alpha_5^*Y_{t-1}^+ - \alpha_6^*Y_{t-1}^-)\eta_t,
\end{equation}
where $\balpha^*= (\alpha_0^*, \alpha_1^*,\ldots,  \alpha_6^*)^\T$ and $\{\eta_t \}_{t \geq 1}$ are \textit{i.i.d.}~standard normal.
It can be shown that the conditional quantile and the expected shortfall functions of $Y_t$ follow the asymmetric slope model given in Example \ref{ASExp}. Especially, for the expected shortfall component, we have $e_t(\bta^*) = \theta_0^* + \theta_1^*e_{t-1}(\bta^*) + \theta_2^*Y_{t-1}^+ - \theta_3^*Y_{t-1}^- + \theta_4^*Y_{t-2}^+ - \theta_5^*Y_{t-2}^-$,
where $\bta^* = \{\alpha_0^* + (1 - \alpha_1^*)\alpha_4^*\ES_\tau(\eta_t), \alpha_1^*, \alpha_2^* + \alpha_5^*\ES_\tau(\eta_t), \alpha_3^* + \alpha_6^*\ES_\tau(\eta_t), -\alpha_1^*\alpha_5^*\ES_\tau(\eta_t), -\alpha_1^*\alpha_6^*\ES_\tau(\eta_t)\}^\T.$
\item[(ii)] The second data generating process is based on the symmetric absolute value model in Example \ref{SAVExp}.
However, $\{Y_t \}_{t\geq 1}$ generated by the following process does not belong to the location-scale family model introduced in~\eqref{full_model_special}.
Let
\begin{equation}\label{eq:global.model}
Y_t = q_t\left\{\btt(U_t) \right\} = \beta_0(U_t) + \beta_1^*q_{t-1}\left\{ \btt(U_t)\right\} + \beta_2(U_t)|Y_{t-1}|,
\end{equation}
where $\{U_t \}_{t\geq 1}$ is a sequence of \textit{i.i.d.}~$\mathrm{Uniform}(0,1)$ random variables.
The vector $\btt(u) = \{\beta_0(u), \beta_1^*, \beta_2(u) \}^\T$ satisfies that given the history $\Hc_t$, $q_t\{\btt(u) \}$ is strictly increasing with respect to $u \in (0,1)$.
For a fixed quantile level $\tau$, we have $e_t(\bta^*) = \theta_0^* + \theta_1^*e_{t-1}(\bta^*) + \theta_2^*|Y_{t-1}|$ with $\bta^* = \{\int_0^\tau \beta_0(u)du/\tau, \beta_1^* ,\int_0^\tau \beta_2(u)du/\tau \}^\T$.
In this section, we set the coefficient functions to $\beta_0(u) = 3F^{-1}_{\text{Beta}}(u) - 1.5$, $\beta_1^* = 0.3$, and $\beta_2(u) = F^{-1}_{\text{Unif}}(u)$, where $F^{-1}_{\text{Beta}}(\cdot)$ and $F^{-1}_{\text{Unif}}(\cdot)$ denote the quantile functions of a $\text{Beta}(2,2)$ distribution and a $\text{Uniform}(-0.8,0.8)$ distribution, respectively.
\end{enumerate}

In Section~\ref{S5.3}, we compare the finite-sample performance of our proposed two-step expected shortfall estimator~\eqref{es_regression_wls} with two competing joint estimators introduced in \cite{P2019} and \cite{DF2020}, referred to as the ``FZ0'' and ``Emp'' estimators, respectively. Detailed descriptions and implementation procedures for these competing methods are provided in Section~S2 of the online supplementary material.

\subsection{Parameter Estimation and Inference}\label{S5.3}
In this section, we evaluate the finite-sample performance of the estimators described in Section~\ref{S5.1}.
Specifically, for each estimator, we compute the element-wise bias and the length of its corresponding $95\%$ confidence interval, based on Theorem \ref{normality_wls} and the corresponding theoretical results in \cite{P2019} and \cite{DF2020}.
To improve the finite sample performance of our proposed two-step expected shortfall estimators, we consider using random weighting bootstrap to construct confidence intervals \citep{Z2020}.
We defer the implementation details of this bootstrap procedure to Section S2.2 of the online supplementary material.
The average bias and confidence interval width, together with the corresponding standard error of the confidence interval width, summarized over $1000$ replications, are reported in Tables \ref{ASres}--\ref{HeteroRes} at the quantile level $\tau = 0.1$.
Moreover, to mitigate the influence of outlier confidence interval lengths arising from numerical instabilities in the optimization routine, we compute the average confidence interval length after applying $95\%$ winsorization.

\begin{table}[!ht]
    \centering
    \scriptsize
    \begin{tabular}{ccccccccccccc}
         \toprule
& \multicolumn{5}{c}{$T = 5000$} && \multicolumn{5}{c}{$T = 7500$} \\
\cmidrule(lr){2-6} \cmidrule(lr){8-12}
& $\theta_1$ & $\theta_2$ & $\theta_3$ & $\theta_4$ & $\theta_5$ && $\theta_1$ & $\theta_2$ & $\theta_3$ & $\theta_4$ & $\theta_5$ \\
\midrule
$\tau = 0.1$ \\
%Avg Bias(LS) & -0.02 & 0.00 & 0.01 & -0.02 & -0.02 &   & -0.01 & 0.00 & 0.01 & -0.01 & -0.01 \\ 
  Avg Bias (WLS) & -0.01 & 0.01 & 0.01 & -0.02 & -0.02 &   & -0.01 & 0.01 & 0.01 & -0.01 & -0.01 \\ 
  %Avg Bias(S.LS) & 0.00 & -0.00 & 0.00 & 0.01 & 0.01 &   & 0.00 & -0.00 & 0.00 & 0.01 & 0.01 \\ 
  %Avg Bias(S.WLS) & 0.01 & 0.01 & 0.02 & -0.00 & -0.01 &   & 0.01 & 0.01 & 0.01 & 0.01 & 0.00 \\ 
  Avg Bias (FZ0) & -0.11 & -0.01 & -0.01 & -0.10 & -0.10 &   & -0.07 & -0.01 & -0.00 & -0.06 & -0.06 \\ 
  %Avg Bias(Oracle) & -0.42 & 0.01 & -0.02 & -0.33 & -0.33 &   & -0.24 & 0.01 & -0.01 & -0.19 & -0.19 \\ 
  Avg Bias (Emp) & -0.95 & 0.12 & 0.04 & -0.56 & -0.55 &   & -0.69 & 0.06 & 0.01 & -0.47 & -0.45 \\  
\cmidrule(lr){2-6} \cmidrule(lr){8-12}
%CI Length(LS) & 0.36 & 0.21 & 0.26 & 0.40 & 0.49 &   & 0.29 & 0.17 & 0.22 & 0.33 & 0.40 \\ 
 %CI Length(LS) & 0.35 & 0.21 & 0.27 & 0.40 & 0.49 &   & 0.29 & 0.17 & 0.22 & 0.33 & 0.40 \\ 
  CI Length (WLS) & 0.38 & 0.19 & 0.25 & 0.41 & 0.50 &  & 0.29 & 0.15 & 0.21 & 0.32 & 0.40  \\ 
  %CI Length(S.LS) & 0.35 & 0.22 & 0.27 & 0.41 & 0.50 &   & 0.29 & 0.18 & 0.23 & 0.34 & 0.41 \\ 
  %CI Length(S.WLS) & 0.32 & 0.21 & 0.29 & 0.38 & 0.49 &   & 0.26 & 0.16 & 0.22 & 0.30 & 0.38 \\ 
  CI Length (FZ0) & 0.39 & 0.22 & 0.31 & 0.44 & 0.55 &   & 0.31 & 0.18 & 0.25 & 0.35 & 0.44 \\ 
 % CI Length (Oracle) & 0.43 & 0.19 & 0.26 & 0.44 & 0.51 &   & 0.32 & 0.16 & 0.21 & 0.33 & 0.40 \\ 
  CI Length (Emp) & 0.55 & 0.25 & 0.37 & 0.54 & 0.63 &   & 0.43 & 0.17 & 0.25 & 0.41 & 0.48 \\ 
\cmidrule(lr){2-6} \cmidrule(lr){8-12}
%Std Err (LS) & 0.38 & 0.13 & 0.18 & 0.28 & 0.33 &   & 0.24 & 0.09 & 0.13 & 0.19 & 0.22 \\ 
 % Std Err (LS) & 0.26 & 0.09 & 0.14 & 0.20 & 0.24 &   & 0.17 & 0.06 & 0.10 & 0.14 & 0.16 \\ 
 ($100\times$) Std Err (WLS)& 0.55 & 0.08 & 0.12 & 0.46 & 0.52 &  & 0.37 & 0.03 & 0.06 & 0.32 & 0.36  \\ 
  %Std Err(S.LS) & 0.28 & 0.09 & 0.15 & 0.21 & 0.26 &   & 0.18 & 0.06 & 0.11 & 0.14 & 0.17 \\ 
  %Std Err (S.WLS) & 0.30 & 0.12 & 0.23 & 0.31 & 0.43 &   & 0.17 & 0.03 & 0.08 & 0.15 & 0.18 \\ 
 ($100\times$) Std Err (FZ0) & 0.40 & 0.05 & 0.09 & 0.31 & 0.33 &   & 0.23 & 0.03 & 0.05 & 0.18 & 0.19 \\ 
  %Std Err(Oracle) & 0.95 & 0.04 & 0.08 & 0.66 & 0.64 &   & 0.54 & 0.03 & 0.06 & 0.36 & 0.33 \\ 
  ($100\times$) Std Err (Emp) & 1.37 & 0.25 & 0.49 & 0.95 & 1.05 &   & 1.01 & 0.06 & 0.17 & 0.72 & 0.74 \\ 
\cmidrule(lr){2-6} \cmidrule(lr){8-12}
 %Coverage(LS) & 0.90 & 0.94 & 0.92 & 0.91 & 0.92 &   & 0.93 & 0.94 & 0.93 & 0.94 & 0.94 \\ 
  Coverage (WLS)  & 0.93 & 0.93 & 0.93 & 0.94 & 0.93 &  & 0.93 & 0.94 & 0.93 & 0.94 & 0.95 \\ 
  %Coverage(S.LS) & 0.84 & 0.94 & 0.91 & 0.87 & 0.88 &   & 0.90 & 0.94 & 0.92 & 0.92 & 0.91 \\ 
  %Coverage(S.WLS) & 0.82 & 0.93 & 0.93 & 0.88 & 0.88 &   & 0.88 & 0.94 & 0.93 & 0.92 & 0.93 \\ 
  Coverage (FZ0) & 0.84 & 0.92 & 0.92 & 0.85 & 0.86 &   & 0.87 & 0.93 & 0.93 & 0.88 & 0.89 \\ 
 % Coverage (Oracle) & 0.65 & 0.90 & 0.86 & 0.67 & 0.67 &   & 0.78 & 0.93 & 0.88 & 0.78 & 0.79 \\ 
  Coverage (Emp) & 0.26 & 0.61 & 0.67 & 0.40 & 0.42 &   & 0.45 & 0.73 & 0.73 & 0.50 & 0.50 \\ 
\bottomrule
    \end{tabular}
    \caption{Simulation results for the asymmetric slope model \eqref{AS_simu} with $\balpha^* = (0.45, 0.65, -0.25,-0.35, 0.8, 0.35, 0.25)^\T$ are reported above. These include the average element-wise bias, the mean and empirical standard errors of the confidence interval lengths, and the coverage probabilities for expected shortfall regression estimators from three different methods: $\hbta_T^W$ from \eqref{es_regression_wls} (WLS), the FZ0 minimization approach proposed in \cite{P2019} (FZ0), the pseudo-efficient joint loss approach proposed in \cite{DF2020} with plug-in estimated nuisance components (Emp).
The mean and standard errors of confidence interval lengths are computed after $95\%$ winsorization.}
\label{ASres}
\end{table}

%From Table \ref{GARCHRes}, we see that under the data generating process in~\eqref{GARCH_simu}, the average bias for all seven estimators approach zero as we increase the sample size.   % yield consistent estimates of $\bta^*$, as indicated by the average bias values approaching zero with increasing sample sizes.
%The empirical coverage probabilities for all estimators are close to the nominal $95\%$ level, supporting the validity of the asymptotic properties derived in Theorems \ref{normality}--\ref{thm:sieve.esreg.normality}. 
%Moreover, confidence interval lengths for the weighted two-step estimator and the joint estimators proposed in \cite{P2019} and \cite{DF2020} have nearly the same performance. 
%Such similarity is due to their asymptotic equivalence, as illustrated in Theorem \ref{normality_wls} and Theorem 2 of \cite{P2019}.
%Specifically, under the data generating process in~\eqref{GARCH_simu}, these estimators can be shown to achieve a broader class of $Z$-estimation efficiency, as shown in Theorem 5.2 of \cite{DF2020}, with only minor modifications to their original proof.
%Finally, we observe that although the sieve-based estimators are asymptotically equivalent to their two-step counterparts, as shown in Theorems \ref{thm:sieve.esreg.normality.ols}--\ref{thm:sieve.esreg.normality}, slight differences in finite-sample performance remain in the empirical performance. These discrepancies are due to the approximation errors introduced by the sieve quantile estimation, which will shrink to zero as the sample size increases.

From Table \ref{ASres}, we find that under the data-generating process in \eqref{AS_simu}, 
the average bias for our proposed estimators $\hbta_T^W$ approaches zero as we increase the sample size.   % yield consistent estimates of $\bta^*$, as indicated by the average bias values approaching zero with increasing sample sizes.
The empirical coverage probability for our proposed WLS estimator is close to the nominal $95\%$ level as the sample size increases, supporting the validity of the asymptotic properties derived in Theorem~\ref{normality_wls}.
For the two joint estimators FZ0 and Emp proposed in \cite{P2019} and \cite{DF2020}, respectively, while they are asymptotically normal, in the complex model setting \eqref{AS_simu}, the joint estimation approach requires solving a more challenging nonconvex optimization problem and estimating more nuisance quantities (see details in Section~S2 of the online supplementary material), which introduce additional numerical instabilities. As a result, their finite-sample performance is worse than that of the two-step estimation approach.
Finally, we observe from Table \ref{ASres} that the proposed weighted two-step estimator has the highest efficiency among all estimators,
as it has the smallest $95\%$ confidence interval length.
This is due to the use of the optimal weighting function in Corollary \ref{ARE}.
%Finally, we observe that although the sieve-based estimators are asymptotically equivalent to their two-step counterparts, as shown in Theorems \ref{thm:sieve.esreg.normality.ols}--\ref{thm:sieve.esreg.normality}, slight differences in finite-sample performance remain in the empirical performance. These discrepancies are due to the approximation error introduced by the sieve quantile estimation, which will shrink to zero as the sample size increases.

\begin{table}[!htbp]
\smaller
    \centering
    \begin{tabular}{cccccccccc}
         \toprule
& \multicolumn{2}{c}{$T = 2500$} & & \multicolumn{2}{c}{$T = 5000$} & & \multicolumn{2}{c}{$T = 7500$} \\
\cmidrule(lr){2-3} \cmidrule(lr){5-6} \cmidrule(lr){8-9}
& $\theta_1$ & $\theta_2$ & & $\theta_1$ & $\theta_2$ & & $\theta_1$ & $\theta_2$ \\
\midrule
$\tau = 0.1$ \\
Avg Bias (WLS) & -0.00 & 0.00 &   & -0.00 & 0.00 &   & -0.00 & 0.00 \\ 
  Avg Bias (FZ0) & 0.00 & -0.00 &   & -0.00 & -0.00 &   & 0.00 & -0.00 \\ 
  Avg Bias (Emp) & 0.00 & -0.00 &   & -0.00 & -0.00 &   & 0.00 & -0.00 \\
  Avg Bias (Optimal) & 0.00 & -0.00 &   & -0.00 & -0.00 &   & 0.00 & -0.00 \\
\cmidrule(lr){2-3} \cmidrule(lr){5-6} \cmidrule(lr){8-9}
   CI Length (WLS) & 0.12 & 0.12 &   & 0.09 & 0.09 &   & 0.07 & 0.07 \\ 
  CI Length (FZ0) & 0.13 & 0.13 &   & 0.09 & 0.09 &   & 0.07 & 0.07 \\ 
  CI Length (Emp) & 0.12 & 0.12 &   & 0.09 & 0.09 &   & 0.07 & 0.07 \\
  CI Length (Optimal) & 0.13 & 0.13 &   & 0.09 & 0.09 &   & 0.07 & 0.07 \\
\cmidrule(lr){2-3} \cmidrule(lr){5-6} \cmidrule(lr){8-9}
($100\times$) Std Err (WLS) & 0.05 & 0.03 &   & 0.02 & 0.01 &   & 0.02 & 0.01 \\ 
 ($100\times$) Std Err (FZ0) & 0.04 & 0.03 &   & 0.02 & 0.01 &   & 0.01 & 0.01 \\ 
($100\times$)  Std Err (Emp) & 0.05 & 0.03 &   & 0.02 & 0.01 &   & 0.01 & 0.01 \\ 
($100\times$)  Std Err (Optimal) & 0.05 & 0.03 &   & 0.02 & 0.01 &   & 0.01 & 0.01 \\ 
\cmidrule(lr){2-3} \cmidrule(lr){5-6} \cmidrule(lr){8-9}
 Coverage (WLS) & 0.93 & 0.93 &   & 0.94 & 0.95 &   & 0.95 & 0.95 \\ 
  Coverage (FZ0) & 0.93 & 0.94 &   & 0.95 & 0.95 &   & 0.95 & 0.95 \\ 
  Coverage (Emp) & 0.93 & 0.94 &   & 0.94 & 0.95 &   & 0.95 & 0.95 \\
  Coverage (Optimal) & 0.93 & 0.94 &   & 0.96 & 0.97 &   & 0.94 & 0.95 \\
\bottomrule 
    \end{tabular}
    \caption{Simulation results for model~\eqref{eq:global.model}. Other details are as in Table \ref{ASres}.}
\label{HeteroRes}
\end{table}

Under model~\eqref{eq:global.model}, the weighting function $1/\{q_t(\btt^*) - e_t(\bta^*) \}^2$ is suboptimal due to the fact that model~\eqref{eq:global.model} does not belong to the location-scale family.
However, motivated by the construction of the optimal weighting function proposed in \cite{B2020}, the conditional variance function of the regression problem~\eqref{es_regression} can be explicitly computed from the data generating process~\eqref{eq:global.model}, such that $\var\{ S_0(Y_t, \bX_t;\btt^*, \bta^*)\mid \Hc_t\} = \int_0^\tau [q_t\{\btt(u) \} - q_t(\btt^*)]^2du/\tau - \{q_t(\btt^*) - e_t(\bta^*) \}^2$.
This quantity can be numerically approximated by solving the quantile regression problem~\eqref{caviar} at multiple quantile levels.
 Consequently, we consider an alternative estimator for this model for comparison, obtained by solving the weighted least squares regression problem in~\eqref{es_regression_wls} with the estimated optimal weighting function.
 We refer to this estimator as the ``Optimal'' estimator and defer the implementation steps to estimate the optimal weighting function to Section~S8 of the online supplementary material.
Our numerical results in Table \ref{HeteroRes} suggest that our proposed weighting function, although suboptimal in the general setting, can have competitive empirical performance relative to commonly used joint estimation approaches and the estimator obtained by optimal weights.

\subsection{Model Misspecification Test}\label{S5.5}
We now assess the empirical performance of the Portmanteau test proposed in Section \ref{LB}.
To this end, we consider a generalized autoregressive conditional heteroskedasticity model motivated by Example \ref{GARCHExp} in Section \ref{S2}, which is also considered in \cite{Z2020}.
We start by introducing the true data generating process under the null hypothesis in \eqref{Port_H0_M}.
Recall from Example \ref{GARCHExp} that
\begin{align}\label{mGARCH0}
Y_t = \sigma_t(\balpha^*)\eta_t,&~~\sigma_t^2(\balpha^*) = \alpha_0^* + \alpha_1^*\sigma_{t-1}^2(\balpha^*) + \alpha_2^*Y_{t-1}^2,
\end{align}
where $\{\eta_t \}_{t \geq 1}$ are \textit{i.i.d.}~standard normal and $\balpha^* = (0.1, 0.1, 0.1)^\T$.
Under \eqref{mGARCH0}, the true conditional quantile and expected shortfall functions are
\begin{equation}
    \label{eq:GARCH_Null}
    q_t(\btt^*) = -\sqrt{\beta_0^* + \beta_1^*q_{t-1}^2(\btt^*) + \beta_2^*Y_{t-1}^2},~~e_t(\bta^*) = -\sqrt{\theta_0^* + \theta_1^*e_{t-1}^2(\bta^*) + \theta_2^*Y_{t-1}^2},
\end{equation}
where $\btt^* = \{\alpha_0^*Q_{\tau}^2(\eta),\alpha_1^*, \alpha_2^*Q_{\tau}^2(\eta) \}^\T$ and $\bta^* =  \{\alpha_0^*\ES_{\tau}^2(\eta),\alpha_1^*, \alpha_2^*\ES_{\tau}^2(\eta) \}^\T$.
Here, we use \eqref{eq:GARCH_Null} as our model under the null hypothesis in \eqref{Port_H0_M}, which corresponds to the scenario in which the conditional quantile and expected shortfall functions are correctly specified.
On the other hand, under the alternative hypothesis, we consider the true data generating process to be
\begin{align}\label{mGARCH}
    Y_t = \sigma_t(\balpha^*)\eta_t,~~\sigma_t^2(\balpha^*) = \alpha_0^* + \alpha_1^*\sigma_{t-1}^2(\balpha^*) + \alpha_2^*Y_{t-1}^2 + cY_{t-2}^2.
\end{align}
We vary the scalar $c \in [0,0.6]$. 
When $c = 0$, model \eqref{mGARCH} reduces to \eqref{mGARCH0}, and the null model is correctly specified. 
For $c > 0$, however, \eqref{mGARCH} deviates from the null data generating process \eqref{mGARCH0}, leading to a misspecification of the null model \eqref{eq:GARCH_Null}.

\begin{comment}
Under the null hypothesis \textcolor{red}{Describe the model first.  the true model does  not change whether it is under null or not. Can you work on this section agian. Very confusing} , where the model is correctly specified, the true data-generating process follows a generalized autoregressive conditional heteroskedasticity process, with parameters $\balpha^* = (0.1, 0.1, 0.1)^\T$.
Recall from Example \ref{GARCHExp}, we have
\begin{align}\label{mGARCH0}
Y_t = \sigma_t(\balpha^*)\eta_t,&~~\sigma_t^2(\balpha^*) = \alpha_0^* + \alpha_1^*\sigma_{t-1}^2(\balpha^*) + \alpha_2^*Y_{t-1}^2,\\
q_t(\btt^*) = -\sqrt{\beta_0^* + \beta_1^*q_{t-1}^2(\btt^*) + \beta_2^*Y_{t-1}^2},&~~e_t(\bta^*) = -\sqrt{\theta_0^* + \theta_1^*e_{t-1}^2(\bta^*) + \theta_2^*Y_{t-1}^2},\notag
\end{align}
where $\btt^* = \{\alpha_0^*Q_{\tau}^2(\eta),\alpha_1^*, \alpha_2^*Q_{\tau}^2(\eta) \}^\T$, $\bta^* =  \{\alpha_0^*\ES_{\tau}^2(\eta),\alpha_1^*, \alpha_2^*\ES_{\tau}^2(\eta) \}^\T$, and $\{\eta_t \}_{t \geq 1}$ \textit{i.i.d.}~standard normal.
Under the alternative hypothesis, we consider true data generating process to be
\begin{align}\label{mGARCH}
    \sigma_t^2(\balpha^*) = \alpha_0^* + \alpha_1^*\sigma_{t-1}^2(\balpha^*) + \alpha_2^*Y_{t-1}^2 + cY_{t-2}^2,
\end{align}
We vary the scalar $c \in [0,0.6]$. 
When $c = 0$, model \eqref{mGARCH} reduces to \eqref{mGARCH0}, and the null model is correctly specified. 
For $c > 0$, however, \eqref{mGARCH} deviates from the null data generating process \eqref{mGARCH0}, which leads to misspecification of the null model.
\end{comment}

We implement the Portmanteau test described in Section \ref{LB} to evaluate whether the null model given in \eqref{mGARCH0} is correctly specified, with $M\in \{4,6,8\}$ in \eqref{Port_H0_M}.
%We set $M=6$ in the hypothesis testing problem in \eqref{Port_H0_M} requires by setting $M = 6$.
We start by assuming that the null hypothesis is true and generate $T = \{2500, 5000, 7500\}$ observations using \eqref{mGARCH0}.
The empirical type I error rate is calculated as the rejection frequency across $1000$ replications of the numerical studies.
The results are summarized in Table~\ref{tab:rejection_frequency}.
We observe that under the null hypothesis, the Portmanteau test controls the Type I error rate around the nominal level of $5\%$ across different choices of $M$.

\begin{table}[!ht]
\centering
\caption{Empirical Type I error rates for different choices of $M$ and $T$ under~\eqref{mGARCH0}.}
\label{tab:rejection_frequency}
\begin{tabular}{c|ccc}
\hline
 & $T = 2500$ & $T = 5000$ & $T = 7500$ \\
\hline
$M = 4$  & 0.056 & 0.041 & 0.043 \\
$M = 6$  & 0.054 & 0.043 & 0.040 \\
$M = 8$  & 0.051 & 0.048 & 0.051 \\
\hline
\end{tabular}
\end{table}
%\textcolor{red}{Remove table and just say it in text? Seems that the table has too few information. }
%\begin{table}[!htbp]
%    \centering
%    \begin{tabular}{cccc}
%    \toprule
%  &  $T = 2500$ & $T = 5000$ & $T = 7500$ \\
%  \midrule
    %$\tau = 0.05$     &  0.064 & 0.056 & 0.054 \\
%    $\tau = 0.1$    &  0.058 & 0.046 & 0.044 \\
    %\bottomrule
    %\end{tabular}
    %\caption{The empirical type I error rates for the hypothesis testing problem \eqref{Port_H0_M}.}
    %\label{BTsimu}
%\end{table}

\begin{figure}[!h]
\centering
  \includegraphics[width=0.4\linewidth]{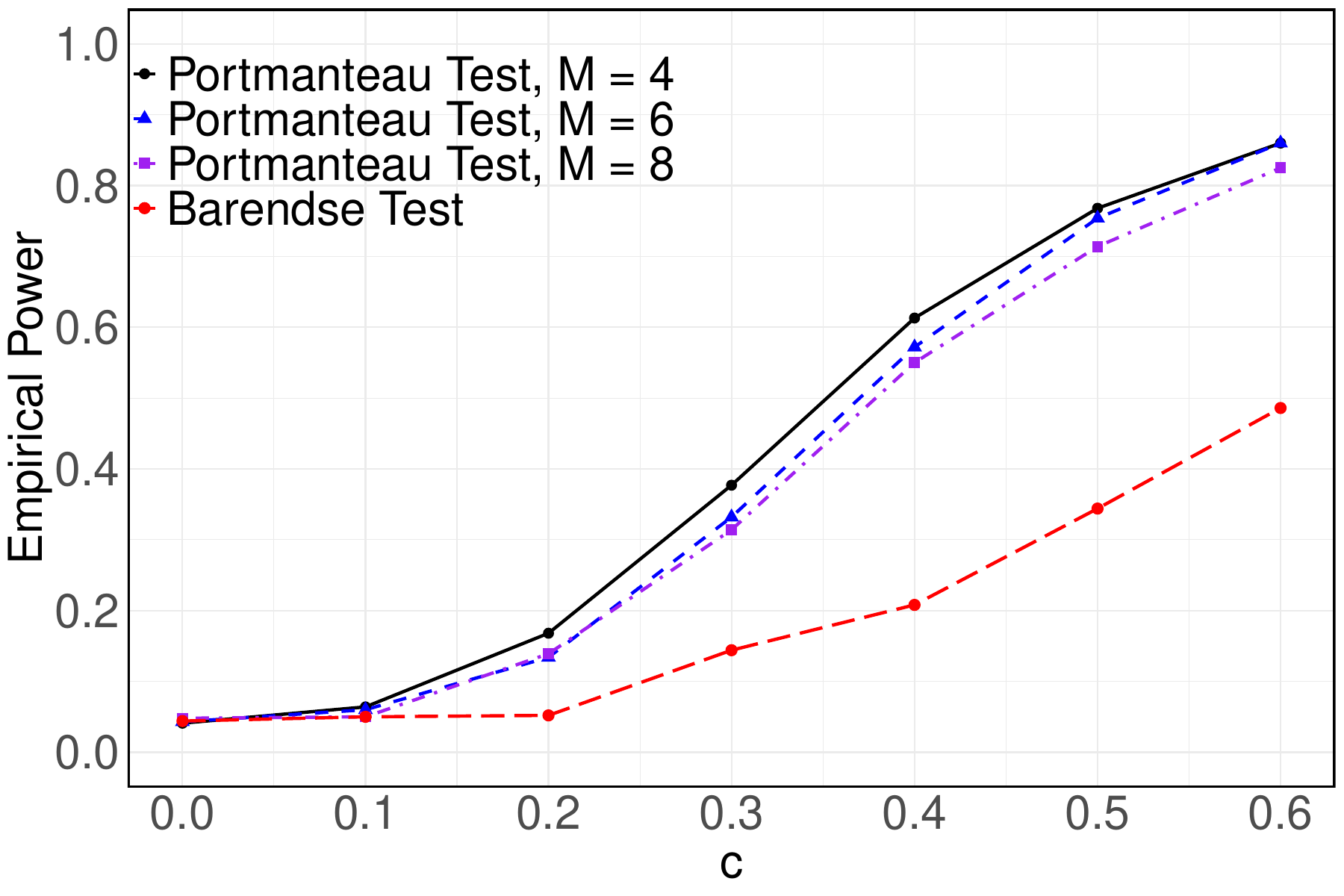}
  \label{garchpower_2}
  %}
  
\caption{The empirical power curve for model misspecification tests with $T = 5000$, $\tau = 0.1$ and $M \in \{4, 6,8 \}$.}
\label{garchpower}
\end{figure}

Next, we evaluate the statistical power of the Portmanteau test.
For comparison, we consider another model diagnostics approach proposed by \cite{BK2023}.
Here, we apply the matrix $T^{(4)}$ given in \cite{BK2023} as the test matrix and follow their proposed approach to model testing.
Under the alternative hypothesis, we set $T = 5000$ for our proposed Portmanteau test.
For the method given in \cite{BK2023}, $T = 5000$ samples are used for in-sample estimation and $50\sqrt{T}$ samples are used for out-of-sample backtesting.
The samples are generated using the alternative model \eqref{mGARCH}, by varying the scalar $c \in \{0, 0.1, 0.2, 0.3, 0.4, 0.5, 0.6\}$. 
The empirical powers of the Portmanteau test and the test proposed in \cite{BK2023} are calculated as the respective null hypothesis rejection frequencies in \eqref{Port_H0_M} and (4) in \cite{BK2023}, respectively. 
Our empirical power curves are shown in Figure \ref{garchpower}. 
As $c$ increases, the data-generating process \eqref{mGARCH} increasingly diverges from the null model \eqref{mGARCH0}, causing the true $Q_\tau(Y_t\mid \Hc_t)$ and $\ES_\tau(Y_t\mid \Hc_t)$ to deviate significantly from the null model \eqref{eq:GARCH_Null}. 
This growing distortion leads to higher rejection frequencies. 
Furthermore, we observe that the power curve for the Portmanteau test consistently dominates the approach in \cite{BK2023}, indicating that our proposed test is empirically more powerful.
The overall patterns of the empirical power curves of the Portmanteau test are highly consistent with different choices of $M$.
Moreover, we include additional numerical experiments on a broader range of model-misspecification settings in Section S9 of the online supplementary material.

\section{Financial Risk Inference and Prediction}\label{S6}

We implement the proposed method to investigate the dynamic behavior of financial risk during periods of global financial crisis.  To this end, we analyze stock index data, specifically the Hang Seng Index and the Dow Jones Industrial Average, from January 2, 2000, to December 31, 2020.
Here, our response variable $Y_t$ is the daily returns of stock indices, which is calculated as $100$ times the log differences of daily closing prices in US dollars.
The covariates $\bm{X}_t$ are set to the lagged values of $Y_t$.

%To investigate the dynamic behavior of financial risk during periods of global financial crisis and to conduct corresponding predictions

We consider two different scenarios: the global financial crisis triggered by the coronavirus pandemic in early 2020, and the real estate-driven financial crisis in 2008.
For the first scenario, we use the data from January 2, 2000 to December 31, 2019 as the training set %, with $T = 4294$ and $T=5031$ 
for the Hang Seng Index and the Dow Jones Industrial Average, respectively, to estimate the model parameters.  We then treat the remaining observations from January 2, 2020, to June 30, 2020 as the test set for performing one-day-ahead predictions of the expected shortfall.
For the second scenario, the model is trained using data from January 3, 2000, to May 30, 2008, and tested on data from June 2, 2008, to December 31, 2008.
Section \ref{S6.1} reports the results for in-sample parameter estimation based on the full in-sample dataset from January 2000 to December 2019. 
Section \ref{S6.2} gives the out-of-sample forecasting results.

%We also consider the prediction of expected shortfall for the year 2008, focusing on the period of the global financial crisis. 
%The model is trained using data from January 3, 2000, to May 30, 2008, and tested on data from June 2, 2008, to December 31, 2008.
%Section \ref{S6.1} reports the in-sample parameter estimation and Portmanteau test results, and section \ref{S6.2} performs the out-of-sample forecasting results.

%In this section, we implement our proposed methodology on real data.

%The data we use is sourced from the stock price database of the Hang Seng Index and Dow Jones Industrial Average.
%We consider stock index data from January 3, 2000, to December 31, 2020.

\subsection{Parameter Estimation and Model Diagnostics}\label{S6.1} 

We begin by determining suitable functional forms for $q_t(\btt)$ and $e_t(\bta)$ in the model \eqref{full_model_special}.
We assume that $q_t(\btt)$ and $e_t(\bta)$ share the same functional form, and consider five candidate models to describe the joint autoregressive dynamics patterns of quantile and expected shortfall.
The models are listed as follows:
\begin{itemize}
    \item[(i)] Lag One Symmetric Absolute Value (SAV1): $e_t(\bta) = \theta_0 + \theta_1e_{t-1}(\bta) + \theta_2|Y_{t-1}|$.
    \item[(ii)] Lag One Asymmetric Slope (AS1): $e_t(\bta) = \theta_0 + \theta_1e_{t-1}(\bta) + \theta_2Y_{t-1}^+ - \theta_3Y_{t-1}^-$.
    \item[(iii)] GARCH: $e_t(\bta) = -\sqrt{\theta_0 + \theta_1e_{t-1}^2(\bta) + \theta_2Y_{t-1}^2}.$
    \item[(iv)] SAV2: $e_t(\bta) = \theta_0 + \theta_1e_{t-1}(\bta) + \theta_2|Y_{t-1}| + \theta_3|Y_{t-2}|.$
    \item[(v)] AS2: $e_t(\bta) = \theta_0 + \theta_1e_{t-1}(\bta) + \theta_2Y_{t-1}^+ - \theta_3Y_{t-1}^- + \theta_4Y_{t-2}^+ - \theta_5Y_{t-2}^-.$
\end{itemize}

We perform the Portmanteau test proposed in Section \ref{LB} for each of the five candidate models for model diagnostics.
We set the lag level $M=10$ for the Portmanteau test, which corresponds to the time duration of two trading weeks.
The test results indicate that the lag one and two symmetric absolute value models are suitable choices for the Hang Seng Index data, with $p$-values $0.28$ and $0.61$, respectively. 
On the other hand, the lag two symmetric and asymmetric slope models serve as more appropriate models to fit Dow Jones Industrial Average data, with $p$-values of $0.06$ for both models.
Based on the above results, we apply the weighted expected shortfall estimator proposed in \eqref{es_regression_wls} to estimate the parameter $\bta$ for the Hang Seng Index and Dow Jones Industrial Average data, respectively.
The results for $\tau = 0.05$ are summarized in Table \ref{tab:combined_theta_est}.

\begin{table}[!h]
\footnotesize
\centering
\begin{tabular}{lcccccccc}
\toprule
& \multicolumn{4}{c}{Hang Seng Index} & \multicolumn{4}{c}{Dow Jones Industrial Average} \\
& \multicolumn{2}{c}{SAV1} & \multicolumn{2}{c}{SAV2} & \multicolumn{2}{c}{SAV2} & \multicolumn{2}{c}{AS2} \\
$\theta_j$ & Est Value & Std Err & Est Value & Std Err & Est Value & Std Err & Est Value & Std Err \\
\midrule
$\theta_0$ & \textbf{-0.09} & 0.02 & \textbf{-0.09} & 0.03 & \textbf{-0.12} & 0.04 & \textbf{-0.10} & 0.03 \\
$\theta_1$ & \textbf{0.90} & 0.02 & \textbf{0.90} & 0.02 & \textbf{0.83} & 0.04 & \textbf{0.87} & 0.04 \\
$\theta_2$ & \textbf{-0.20} & 0.05 & -0.16 & 0.09 & -0.25 & 0.13 & -0.06 & 0.17 \\
$\theta_3$ & --- & --- & -0.04 & 0.09 & -0.10 & 0.14 & \textbf{-0.44} & 0.19 \\
$\theta_4$ & --- & --- & --- & --- & --- & --- & -0.01 & 0.12 \\
$\theta_5$ & --- & --- & --- & --- & --- & --- & 0.01 & 0.18 \\
\bottomrule
\end{tabular}
\caption{Estimated values and standard errors for $\theta_j$ in $e_t(\bta)$ across different models and indices at $\tau = 0.05$. Bold entries indicate significance at the 5\% confidence level.}
\label{tab:combined_theta_est}
\end{table}

From Table \ref{tab:combined_theta_est}, we see that the estimated values for $\theta_1$, the coefficient of the autoregressive term, are always significant across different models and stock indices.
Our results align with similar conclusions in the context of autoregressive conditional quantile regression \cite{E2004}, thus supporting the phenomenon of clustering patterns in conditional expected shortfall.
Next, we find that the expected shortfall dynamic patterns are different between the Hang Seng Index and the Dow Jones Industrial Average.
The symmetric absolute value models fit the Hang Seng Index data well, indicating that the previous positive return and negative return have similar influences on future risk.
On the other hand, we find that the asymmetric slope model quantifies the differential impact of previous positive and negative returns for the Dow Jones Industrial Average, as $\theta_3$, the coefficient for $-Y_{t-1}^-$, is significantly different from $\theta_2$, the coefficient of $Y_{t-1}^+$.
This indicates a high sensitivity of the financial market to the negative return of the previous day.
This asymmetric influence between previous positive and negative returns suggests strong evidence of the leverage effect, which stems from the empirical fact that losses have a greater influence on future risks than gains \citep{C1982,Yu2005}.

In practice, when prior knowledge of the model structure is not available, we recommend that users apply the workflow described in this section for analyzing financial time series data. Specifically, we first implement the Portmanteau test on the time series data over a finite set of candidate joint quantile-expected shortfall models. We then select the models that pass the Portmanteau test as plausible working models and use them to fit the data.

%We next conduct model diagnostics on the two working models, respectively.
%We obtain the expected shortfall regression residuals by using results given in Tables \ref{HSI_res}--\ref{DJ_res} and conduct Ljung-Box test on the residual sequences.
%The lag level of the Ljung-Box test is fixed to be five, which corresponds to a whole trading week.
%The testing results are given in Table \ref{LBdata}.
%For the Hang Seng Index return sequence, both lag two asymmetric slope model and generalized autoregressive conditional heteroskedasticity model can fit the data well for lower quantile levels, since the Ljung-Box test failed to reject the null hypothesis.
%For Dow Jones Industrial Average data, only the residuals under the lag two asymmetric slope model shows no significant serial autocorrelation for $\tau \in \{0.025, 0.05\}$,
%which suggests us to pick this model as an approporiate working model over the generalized autoregressive conditional heteroskedasticity model for lower quantile levels.

\subsection{Out-of-Sample Expected Shortfall Prediction}\label{S6.2}
In this section, we evaluate the predictive performance of our proposed dynamic autoregressive expected shortfall model during periods of significant financial turbulence. 
We focus on two distinct testing periods, as detailed in the beginning of Section \ref{S6}.
%The first spans from June 30, 2008, to December 31, 2008, capturing the apex of the global financial crisis driven by the subprime mortgage collapse. 
%The second period starts from January 2, 2020, to June 30, 2020, during which the global markets faced severe volatility induced by the coronavirus disease 2019 pandemic.
%The risk level in this section is fixed to be $\tau = 0.05$.

Following the training-testing split described in Section \ref{S6} and the results given in Section \ref{S6.1}, we select the lag two symmetric absolute value model and lag two asymmetric slope model to make expected shortfall predictions for the Hang Seng index and Dow Jones Industrial Average data, respectively.
In addition to these two models, we also consider a rolling window approach for predicting expected shortfall. Specifically, 
for a given time point $ t $ and a predetermined window size $ h $, the conditional $\tau$-level expected shortfall of $ Y_t $ is estimated as the empirical marginal $\tau$-level expected shortfall of the series $\{Y_{t-h}, Y_{t-h+1}, \ldots, Y_{t-1}\}$. In this analysis, we set the window size $h$ to be $125$ for simplicity.

\begin{figure}[!htbp]
    \centering
    \includegraphics[width=1\linewidth]{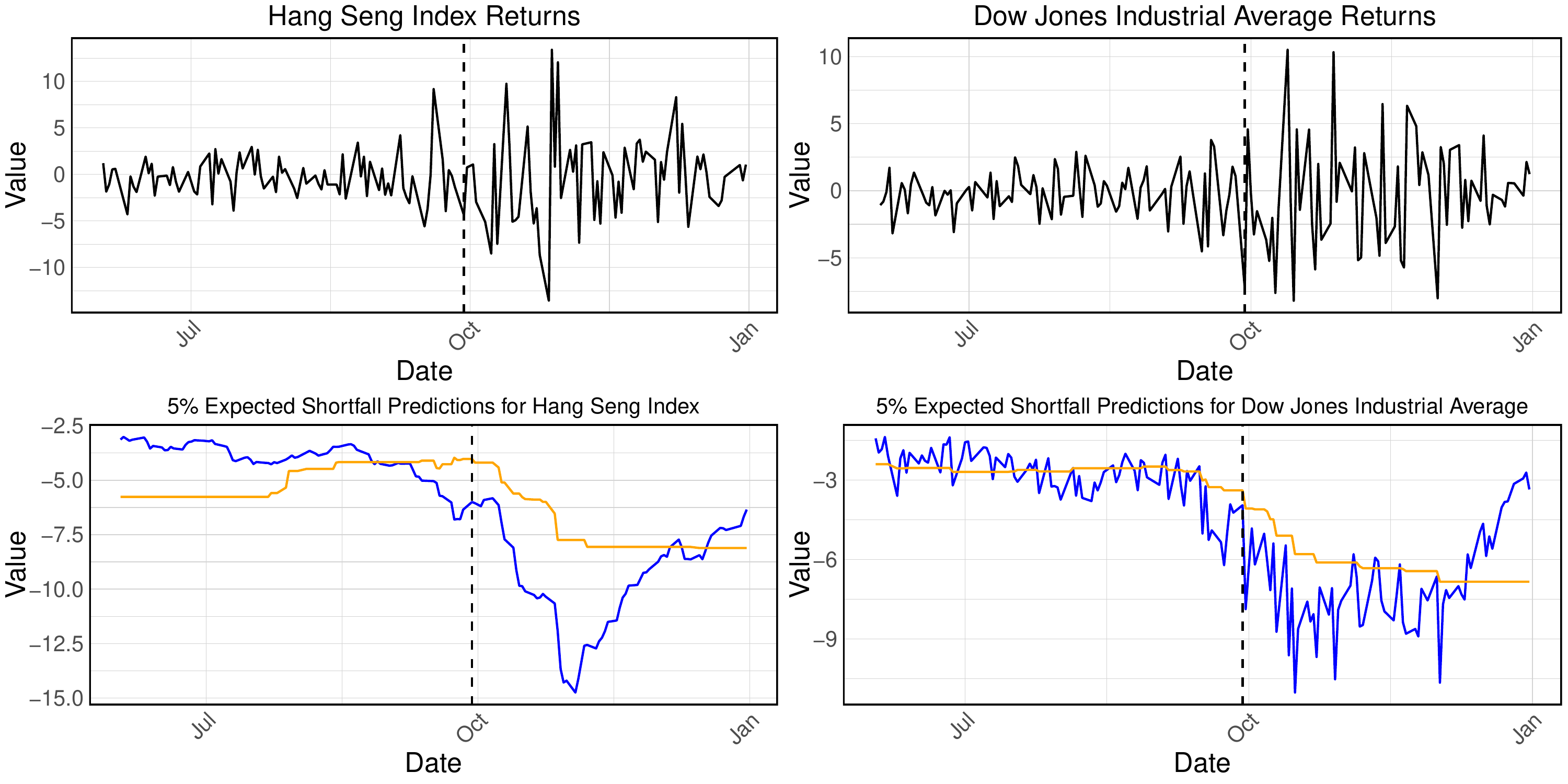}
    \caption{The $5\%$ level expected shortfall forecasting results for different approaches: rolling window (orange curve), lag two symmetric absolute value and asymmetric slope models (blue curves). The upper panel includes the plots for the observed Hang Seng Index and Dow Jones Industrial Average return series in the year 2008. The lower panel plots the one-step expected shortfall predicted values for the aforementioned three models. The dashed vertical black line denotes the date of the largest single-day point drop in the history of the Dow Jones Industrial Average, which occurred on September 29, 2008.}
    \label{ESPred_Res_2008}
\end{figure}

\begin{figure}[!ht]
    \centering
    \includegraphics[width=1\linewidth]{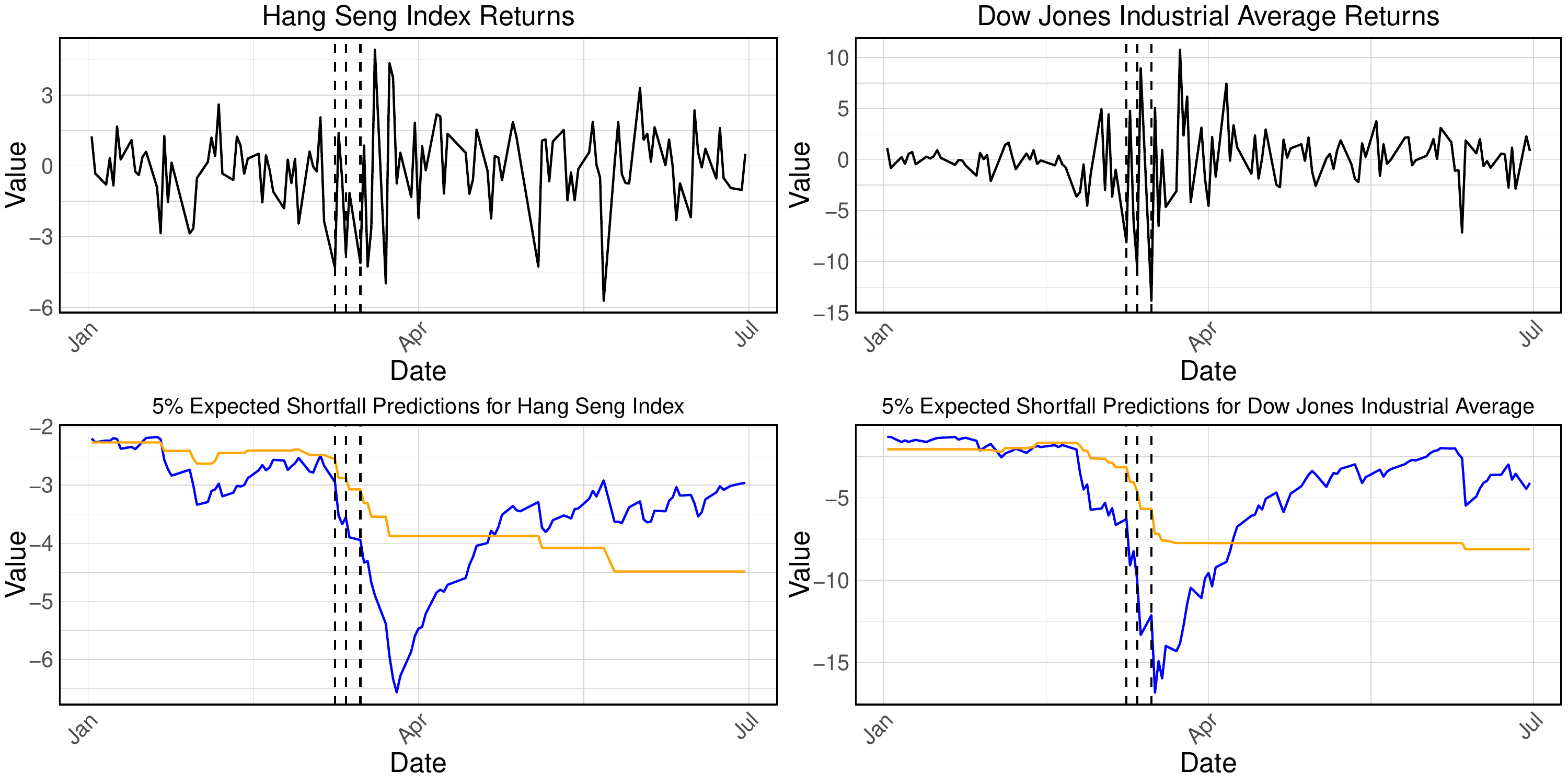}
    \caption{The $5\%$ level expected shortfall forecasting results for different approaches: rolling window (orange curve), lag two symmetric absolute value and asymmetric slope models (blue curves). The upper panel contains plots for the observed Hang Seng Index and Dow Jones Industrial Average return series in the year 2020. The lower panel plots the one-step expected shortfall predicted values for the aforementioned three models. The dashed vertical black lines denote the ``Black Monday I'', ``Black Thursday'' and ``Black Monday II'' stock market crashes on March 9, 2020, March 12, 2020 and March 16, 2020.}
    \label{ESPred_Res_2020}
\end{figure}

The forecast results are presented in Figures \ref{ESPred_Res_2008}--\ref{ESPred_Res_2020}. 
Our initial observation is that the rolling window-based approach provides only a rough prediction and cannot reveal the complex dynamic pattern of financial risk.
Conversely, the predictions made by the lag two asymmetric slope and symmetric absolute value models can all reveal more complex dynamic patterns of the expected shortfall throughout the year than the rolling window approach.
Both models exhibit very similar trends in predicting one-day-ahead conditional expected shortfall. 
The prediction results for 2008, as shown in Figure \ref{ESPred_Res_2008}, indicate a significant decline in the expected shortfall estimates for both the Hang Seng Index and the Dow Jones Industrial Average following September 29, 2008.
This date is particularly important for the Dow Jones Industrial Average, as it recorded the largest single-day point drop in its history.
In 2020, the global stock market experienced three major events due to the pandemic: Black Monday I on March 9, Black Thursday on March 12, and Black Monday II on March 16. 
As illustrated in Figure \ref{ESPred_Res_2020}, the expected shortfall predictions for both models and both stock indices show a similar downward trend during this period.

In summary, both the lag two symmetric absolute value and the asymmetric slope models effectively predict the risk levels of various stock indices. 
Moreover, as discussed in Section \ref{S6.1}, the lag-two asymmetric slope model for Dow Jones data offers the advantage of distinguishing between the differential impacts of past gains and losses on stock returns. 
Due to its greater flexibility, we recommend adopting the lag-two asymmetric slope model as the preferred working model in this example. % as long as the guarantee from Portmanteau test is obtained.

\section{Discussion}\label{S7}
We consider a joint model for conditional quantiles and expected shortfall with autoregressive structures. 
%The proposed model encompasses a wide variety of existing time series models, including autoregressive-moving average models, generalized autoregressive conditional heteroskedasticity models, and quasi-score driven models. 
Existing estimation approaches for such models typically involve minimizing a class of joint loss functions \citep{FZ2016,P2019}. 
Motivated by \cite{B2020}, we propose a weighted two-step estimation scheme to estimate the expected shortfall parameters.
%\textcolor{blue}{
%The key idea of our estimation methodologies is to compute the quantiles in the initial step as nuisance components, subsequently employing a regression approach to estimate the expected shortfalls. 
%Our initial scheme solves the nonlinear quantile regression problem \eqref{caviar} as introduced in \cite{E2004}.
%This approach yields a Portmanteau test as introduced in Section \ref{LB} as a by-product.
%However the non-linear quantile regression problem \eqref{caviar} is hard to optimize.
%To mitigate such computational challenges, our second approach employs a sieved linear model to approximate the autoregressive quantile functions. 
%This adaptation simplifies the optimization problem \eqref{caviar} to a linear quantile regression problem, thereby enhancing computational efficiency.}
We start by using an unweighted least squares approach to estimate the expected shortfall parameters under the model \eqref{full_model}.
We apply weighted least squares regression to improve the asymptotic efficiency of our estimators.
The choice of our weighting function is motivated by the location-scale family time series with \textit{i.i.d.}~innovations.
Under such a model assumption, our proposed weighting functions are optimal and can achieve asymptotic efficiency among the class of joint estimators proposed in \cite{DF2020}.
Moreover, we develop a Portmanteau test based on \eqref{full_model_special} with \textit{i.i.d.}~noise assumption, which exhibits great empirical efficiency.
Both theoretical analysis and numerical simulations demonstrate that our proposed estimator is more efficient than those of \cite{P2019}.

Although the location-scale family time series models with \textit{i.i.d.}~innovations encompass a wide range of commonly used time series models, they remain a specific subclass of the more general model \eqref{full_model}. 
%For instance, if the noise series \(\{\zeta_t\}_{t \geq 1}\) is dependent or heterogeneous, model \eqref{full_model} remains valid, but \eqref{full_model_special} does not. 
In some scenarios, our proposed weighted estimator is suboptimal. % and loses its asymptotic efficiency.  
A potential direction for future research is the development of nonparametric approaches to derive optimal weights without relying on the assumption in \eqref{full_model_special}.
Moreover, it is well known that the least squares step involved in the two-step approach is highly sensitive to heavy-tailed and highly skewed data, which are common in financial asset returns. 
Incorporating a robust loss function into the two-step approach may address this issue and yield a robust estimator \citep{H2023}. 
We will leave the aforementioned discussion for future research.




%\bibliographystyle{elsarticle-harv} 
\bibliography{reference}

\end{document}